\documentclass[manuscript]{aastex}

\usepackage{amssymb}
\usepackage{epsfig}
\usepackage{amsmath}
\usepackage{natbib}
\usepackage{color}

\slugcomment{submitted to PASP}

\shorttitle{Kepler eclipses and phase-curves}
\shortauthors{Angerhausen et al.}

\begin{document}


\title{A comprehensive study of Kepler phase curves and secondary eclipses --  temperatures and albedos of confirmed Kepler giant planets}


\author{Daniel Angerhausen}
\affil{Rensselaer Polytechnic Institute, 110 Eighth Street, Troy, NY 12180; \\ORAU NASA Postdoctoral Fellow, NASA Goddard Space Flight Center, Greenbelt, MD 20771  }

\author{Em DeLarme}
\affil{Rensselaer Polytechnic Institute, 110 Eighth Street, Troy, NY 12180; \\ UCF, Planetary Sciences Group, Department of Physics, Orlando, FL 32816}

\and

\author{Jon A. Morse}
\affil{Rensselaer Polytechnic Institute, 110 Eighth Street, Troy, NY 12180; \\ BoldlyGo Institute, 1370 Broadway, 5th Floor Suite 572, New York, NY 10018}



\begin{abstract}
We present a comprehensive study of phase curves and secondary eclipses in the Kepler data set using all data from 16 quarters that were available in 2013-2014.
Our sample consists of 20 confirmed planets with $R_p > 4 R_e$ , $P < 10d$, $V_{mag} < 15$. Here we derive their temperatures and albedos, with an eye towards constraining models for the formation and evolution of such planets. Where there was overlap our results confirm parameters derived by previous studies, whereas we present new results for Kepler 1b-8b, 12b-15b, 17b, 40b, 41b, 43b, 44b, 76b, 77b, and 412b derived in a consistent manner. We also present lightcurve analyses for Kepler 91b and Kepler 74b, which both show extra dimmings at times other than from the expected primary and secondary eclipses. Corrected for thermal emission we find most of the massive planets from our sample to be low in albedo ($<0.1$) with a few having higher albedo ($>0.1$). 

\end{abstract}


\keywords{planets and satellites: atmospheres
planets and satellites: fundamental parameters
planets and satellites: gaseous planets
planets and satellites: general}



\section{Introduction}

\subsection{The \textsl{Kepler} mission}
Studying extrasolar planets is one of the major frontiers of astronomy today.  The field has transformed from simple identification to comprehensive categorization and characterization
of exoplanets and exoplanetary systems. Analyses of data provided by NASA's \textsl{Kepler}\footnote{The data presented in this paper were obtained from the Mikulski Archive for Space Telescopes (MAST), which is managed by the Space Telescope Science Institute (STScI). STScI is operated by the Association of Universities for Research in Astronomy, Inc., under NASA contract NAS5-26555. Support for MAST for non-HST data is provided by the NASA Science Mission Directorate via grant NNX13AC07G and by other grants and contracts. This paper includes data collected by the \textsl{Kepler} mission. Funding for the \textsl{Kepler} mission is also provided by the NASA Science Mission Directorate.} mission has revolutionized this field by compiling a statistically significant number of transiting planets and planetary candidates (e.g.,
\citealt{2010Sci...327..977B}; \citealt{2011ApJ...736...19B}; \citealt{2013ApJS..204...24B}; \citealt{2014ApJS..210...19B}; \citealt{2015ApJS..217...16R}; \citealt{2015ApJS..217...31M}).

For example, \textsl{Kepler} data allowed researchers to discover Kepler 9b (\citealt{2010Sci...330...51H}), the first multi-planetary system outside our solar system; Kepler 10b (\citealt{2011ApJ...729...27B}), one of  the first confirmed rocky planets outside the solar system; and Kepler 16b (\citealt{2011Sci...333.1602D}), the first circumbinary planet. More recently the \textsl{Kepler} team announced the discovery of potentially habitable worlds in the Kepler 62 (\citealt{2013Sci...340..587B}) and Kepler 69 (\citealt{2013ApJ...768..101B}) systems, and the near-Earth-sized planets Kepler-186f (\citealt{2014Sci...344..277Q}) and Kepler-452b (\citealt{2015AJ....150...56J}) in the habitable zones of their parent stars.

Deeper analyses are possible using the exquisite \textsl{Kepler} data beyond merely detecting exoplanetary systems: researchers are now able to analyze large samples of planetary candidates to pin down occurrence rates such as $\eta_{earth}$ (e.g., \citealt{2012ApJS..201...15H}; \citealt{2013ApJ...767...95D}; \citealt{2013ApJ...766...81F}; \citealt{2013PNAS..11019273P}; \citealt{2015arXiv150604175B}), find non-transiting planets via transit timing variations (\citealt{2011ApJ...743..200B}), perform phase-curve analyses (\citealt{2013ApJ...771...26F}), and may eventually even be able to detect exomoons (\citealt{2013ApJ...770..101K}) and exotrojans (\citealt{2015arXiv150800427H}).

For the close-in, and therefore hot, planets around bright, high-signal host stars in the \textsl{Kepler} data set, we are able to analyze secondary eclipses, i.e., the modulated flux from the star-planet system when the light (reflection and thermal emission)  of the planet disappears during its passage behind the parent star. Differential measurements then help us to characterize physical parameters of the planet such as albedo and temperature.
Measuring such quantities, together with complementary spectroscopic measurements, can provide contraints on models for the formation and atmospheric photochemistry of such close-in planets \citep[e.g.,][]{2010ApJ...717..496L, 2011ApJ...737...15M, 2011ApJ...738...72V}

\begin{deluxetable}{lll}
\tabletypesize{\scriptsize}
\tablecaption{\textsl{Kepler} Quarters used in our analysis\label{tbl:used_quarters}}
\tablewidth{0pt}
\tablehead{
\colhead{KOI} & \colhead{LC Quarters} & \colhead{SC Quarters}}
\startdata
1    &0,1,2,3,4,5,6,7,9,10,11,13,14,15 & 0,1,2,3,4,5,6,7,8,9,10,11,12,13,14,15 \\
2    &0,1,2,3,4,5,6,7,8,9,10,11,12,13,14,15& 0,1,2,3,4,5,6,7,8,9,10,11,12,13,14,15 \\
3    &0,1,2,3,4,5,6,8,9,10,12,13,14& 0,1,2,3,4,5,6,8,9,10,11,12,13,14 \\
7    &0,1,2,3,4,5,6,7,9,10,11,13,15& 0,1,2,3,4,5,6,7,9,10,11,13,14,15 \\
10    &0,1,2,3,4,5,6,7,8,9,10,11,12,13,14,15& 1,2,3,4,5,6,7,8,9,10,11,12,13 \\
13    &0,1,2,3,4,5,6,7,8,9,10,11,12,13,14,15& 1,2,3,6,7,8,9,10,11,12,13,14,15 \\
17    &0,1,2,3,4,5,6,8,9,10,12,13,14& 1,2,3,4,5,6,8,9,10,11,12 \\
18    &0,1,2,3,4,5,6,7,8,9,10,11,12,13,14,15& 1,2,3,4,5,6,7,8,9,10,11,12 \\
20    &0,1,2,3,4,5,6,7,9,10,11,13,14,15& 1,2,3,4,5,6,7,8,9,10,11 \\
97    &0,1,2,3,4,5,6,7,8,9,10,11,12,13,14,15& 2,3,4,5,6,7,8 \\
98    &0,1,2,3,4,5,6,7,8,9,10,11,12,13,14,15& 2,3,4,5,6,7,8,9,10,11,12 \\
127    &1,2,3,4,5,6,7,8,9,10,11,12,13,14,15& 2,3,4,5,6,7 \\
128    &1,2,3,4,5,6,7,8,9,10,11,12,13,14,15& 2,3,4,5,6,7 \\
135    &1,2,3,4,5,6,7,8,9,10,11,12,13,14,15& 2,3,4,5,6,7,9,10,11,12,13,14,15 \\
196    &1,2,3,4,5,6,7,8,9,10,11,12,13,14,15& 3,4,5,6,7 \\
200    &1,2,3,4,5,6,7,8,9,10,11,12,13,14,15& 3,4,5,6,7 \\
202    &1,2,3,4,5,6,7,8,9,10,11,12,13,14,15& 3,4,5,6,7 \\
203    &1,2,3,4,5,6,8,9,10,12,13,14& 3,4,5,6,7,8,9,10,11,12,13,14 \\
204    &1,2,3,4,5,6,7,8,9,10,11,12,13,14,15& 3,4,5,6,7 \\
428    &1,2,3,4,5,6,8,9,10,12,13,14& -- \\
1658    &1,2,3,4,5,7,8,9,11,12,13,15& --\\
2133    &0,1,2,3,4,5,6,7,8,9,10,11,12,13,14,15& --

\enddata
\tablecomments{For our analysis we used all data available prior to data release Q16, August 2013}

\end{deluxetable}

\subsection{Transits and eclipses}

Systems with transiting extrasolar planets can offer two important
observational opportunities for deriving physical parameters of the planets. In \emph{primary
transit} the planet crosses the star. From a broadband
transit-lightcurve, in this case, one can measure the planetary radius
$R_{p}$ in units of the stellar radius $R_{*}$. The depth of the
transit is $\sim(R_{p}/R_{*})^{2}$, which for a Jupiter radius
planet transiting a sun-like star, is of the order of $\sim 1\%$
(e.g., \citealt{2000ApJ...529L..41H}).

If the geometry (inclination, eccentricity) is right the planet also disappears behind its host star in a so-called  \emph{secondary eclipse}.  For a 2000 K hot Jupiter-size planet, the typical flux deficit during
secondary eclipse is $\sim 200$ ppm at $\sim 2 \mu m$ in the near-infrared and even larger at longer wavelengths, but considerably smaller at optical wavelengths (400-900 nm) at which \textsl{Kepler} observes. However, for host stars that are bright enough, \textsl{Kepler}'s outstanding sensitivity provides a direct measure of the planet's disk-averaged day side flux for some of its targets, particularly close-in gas giants -- so called 'Hot Jupiters' and 'Hot Neptunes'.

Observing secondary eclipses combined with planetary phase curves can help us to characterize the planet and its atmosphere. For example, the depth of
the secondary eclipse can constrain the albedo of the
planet, while the timing and width of the secondary
eclipse can help determine its orbital parameters.
Comparing the amplitude of the reflected light in the phase curve with the depth of the secondary eclipse can constrain the day and night side temperatures, and therefore confirm the planetary nature of a candidate that is not self-luminous, and help to understand day to night side heat exchange.
Measuring exoplanet eclipses, phase curves and albedo values yield information about the composition of their atmosphere, and day to night side temperature ratios yield the efficiency of energy transport and presence of possible temperature inversions (for details see, e.g., \citealt{2011ApJ...729...54C}).

Several previous studies have focused on eclipses and phase curves in the \textsl{Kepler} database, either on select samples of objects (e.g., \citealt{2011ApJ...730...50K} 5 objects; \citealt{2012AJ....143...39C} 76 objects; \citealt{2013ApJ...772...51E} 8 objects) or for individual planets or candidates (e.g., \citealt{2012AandA...541A..56M}; \citealt{2013ApJ...764L..22M}).

In the largest sample so far, \citealt{2012AJ....143...39C} modeled  secondary eclipses and phase curves (only via a simple sinusoidal flux term applied to the lightcurves) of a uniform set of Hot Jupiter candidates from \textsl{Kepler} to derive albedos and thermal emission properties, and compared the results with stellar and planetary parameters. While our study is similar in scope to this study, we worked with much more data (15 quarters -- as available in August 2013 -- of data instead of 1), a slightly bigger radius range (4 Earth radii = 0.4 Jupiter radii, instead of 0.5 Jupiter radii as a lower limit), and a longer period range (10 days instead of 5 days). Furthermore we applied a fully physical model with all different phase curve components. Of the \citealt{2012AJ....143...39C} sample of 76 KOIs only 55 were successfully modeled with their methods. Of these 55 systems many still remain unconfirmed planetary candidates or turned out to be false postives (e.g., KOIs 102, 1419, 1459, 1541, 1543; http://exoplanetarchive.ipac.caltech.edu/, August 2015).
For the confirmed planets there is significant overlap in our samples (KOIs 1, 2, 10, 13, 17, 18, 20, 97, 127, 128, 196, 202, 203, 204), but several planets in our study (KOIs 3, 7, 98, 135, 428 and 1658) were not covered by their analysis.

Here we present initial results of a comprehensive and consistent study of secondary eclipses and phase curves using data from quarters 0 through 15 of \textsl{Kepler} lightcurves (see Table \ref{tbl:used_quarters}), that were available at the time of our analysis. In this paper we focus on the 20 confirmed (August 2013) planets in the sample of 489 Kepler Objects of Interest (KOI) with $R_p > 4 R_e$, $P < 10d$, and $V_{mag} < 15$. 
Consistent measurements of exoplanet phase curves not only allows us to break many current degeneracies in modeling the thermal and chemical structure of separate exoplanet atmospheres, but also enables us to compare results across the whole set of analyzed systems in a comprehensive way. Analyses like this will also help us prepare for future ground- and space-based facilities that increase the number of exoplanetary systems and the wavelength range of precision observations.

\section{Data reduction}
\subsection{PyKE data preparation}

For each of our targets, we used up to 16  quarters of the Pre-search Data Conditioning Simple Aperture Photometry (PDCSAP) lightcurves from the \textsl{Kepler} database available in the Mikulski Archive for Space Telescopes (MAST) at the time of our analysis (see Table \ref{tbl:used_quarters}; prior to the release of Q16 in Aug 2013)).  The PDCSAP lightcurves are simple aperture photometry timeseries that have been cotrended in the \textsl{Kepler} pipeline to remove systematics common to multiple targets, using a best-fit of so-called `Cotrending Basis Vectors' (CBVs). The CBVs are essentially the principal components of systematic artifacts for each science target and each operational quarter characterized by quantifying the features most common to hundreds of strategically-selected quiet targets sampled across the detector array (see \citealt{2012PASP..124.1000S}; \citealt{2012PASP..124..985S}). Short-cadence data were used in place of long-cadence data when available and both data sets were combined weighted by their errors. 

We used PyKE (\citealt{2012ascl.soft08004S}), a series of python-based PyRAF recipes, for the individual and target-specific analysis and reduction of \textsl{Kepler} timeseries data. Our first step was to remove long-term variability using the \texttt{kepflatten} task to fit a quadratic polynomial to the baseline parts of the lightcurve for each quarter over time intervals of seven times the length of the published orbital period of each planet (see Table \ref{tbl:used_params}) using a sigma clipping of 2.5. We thus minimize any contamination or over-correction of the actual planetary phasecurves by the polynomial flattening and excluded the occultations.  We then concatenated these flattened curves using the \texttt{kepstitch} routine, which created one long timeseries, containing all used quarters of data (see Table \ref{tbl:used_quarters}). Finally, using the \texttt{kepfold} task, we folded the entire lightcurve by the published orbital periods of each planet to get the phase curve, that was used in the next steps of our data analysis.

\begin{deluxetable}{llcccccccc}
\tabletypesize{\scriptsize}
\tablecaption{System parameters used in phasecurve fits\label{tbl:used_params}}
\tablewidth{0pt}
\tablehead{
\colhead{KOI} & \colhead{reference}& \colhead{period [d]} & \colhead{$T_{eff}$ [K]} & \colhead{$log(g)$} & \colhead{$[Fe/H]$} & \colhead{\textit{u}\tablenotemark{a}} &
\colhead{\textit{y}\tablenotemark{a}} & \colhead{$R_p/R_*$\tablenotemark{b}} & \colhead{$a/R_*$\tablenotemark{b}}}

\startdata
      1&  \cite{2009yCat.5133....0K}   & 2.47061317&5713&      4.14&    -0.14&     0.62&     0.38&0.124&8.445\\
      2&   \cite{2011ApJ...736...19B}   &2.20473537&6577&      4.32&     0.26&     0.54&     0.25&0.075&4.681\\
      3&  \cite{2010ApJ...710.1724B}    &4.88780026&4780&      4.59&     0.31&    0.72&    0.55&0.057&16.68\\
      7&   \cite{2010ApJ...713L.126B}   &3.2136641&5857&      4.25&   0.17&     0.61&     0.36&0.025&6.47\\
     10&  \cite{2010ApJ...724.1108J}    &3.52249913&6213&      4.28&    -0.05&    0.57&     0.30&0.093&7.5\\
     13&  \cite{2009yCat.5133....0K}    &1.7635877&8848&      3.93&    -0.14&     0.46&     0.58&0.077&4.4\\
     17&    \cite{2010ApJ...713L.136D}  &3.23469955&5647&      4.23&     0.34&    0.64&     0.40&0.093&7.53\\
     18&   \cite{2011ApJ...736...19B}   &3.54846566&5816&      4.46&    0.04&     0.61&     0.36&0.078&7.19\\
     20&   \cite{2011ApJS..197....9F}   &4.43796291&5947&      4.17&    0.07&    0.60&     0.34&0.117&8.133\\
     97&    \cite{2010ApJ...713L.140L}  &4.88548917&5933&      3.98&    0.11&     0.60&     0.34&0.082&6.842\\
     98& \cite{2011ApJS..197....3B}     & 6.7901235&6395&      4.11&    0.12&     0.55&     0.29&0.056&7.299\\
    127&  \cite{2013AandA...557A..74G}    &3.57878272&5520&      4.4&  0.20 &     0.63&     0.41&0.096&10.36\\    
    128&   \cite{2009yCat.5133....0K}   &4.94278327&5718&      4.18&     0.36&     0.64&     0.39&0.101&10.34\\
    135&  \cite{2012AandA...538A..96B}    &3.02409489&6041&      4.26&     0.33&    0.59&     0.33&	0.0805 & 4.681\\
    196&  \cite{2011AandA...536A..70S}    &1.85555773&5660&      4.44&     -0.09&   0.62&     0.39&0.096&5.99\\
    202&    \cite{2014arXiv1401.6811D}  &1.72086037&5750&      4.3&   0.27 &     0.61&     0.38&0.099&5.22\\
     
    203&   \cite{2012AandA...538A..96B}  & 1.48571127&5781&      4.53&    0.26&     0.61&     0.37&0.129&5.67\\
    204&  \cite{2012AandA...538A..96B}   & 3.246732&5757&      4.15&    0.26&     0.61&     0.37&0.075&8.5\\
    428&    \cite{2011AandA...528A..63S}   &6.8731791&6510&      3.94&    0.1&     0.53&     0.26& 0.059 &6.275\\
   1658&   \cite{2013ApJ...771...26F}   &1.54492977&6300&      4.2&   -0.1 &     0.53&     0.28& 0.085 &4.61

\enddata
\tablenotetext{a}{Limb- and gravity-darkening parameters derived from \cite{2011AandA...529A..75C}.}
\tablenotetext{b}{Where needed parameters were given in units of stellar mass $M_*$ and radius $R_*$ and therefore independent of these values. }

\end{deluxetable}

\subsection{Phase curve model}

We removed the primary transit signature as a first step in modeling the phase-folded lightcurves. The remaining normalized, out-of-transit phase curve $F_{tot}$ was then modeled as
\begin{eqnarray}
  F_{tot} &=& f_0 + F_e + F_d + F_p + F_{ecl} 
\end{eqnarray}
where $F_{tot}$ is the sum of the stellar baseline $f_0$ (1 for normalized data) and the following four contributions to the lightcurve as a function of phase $\phi$, with $\phi \in [0,1] $, the primary transit at $\phi=0$ and the secondary eclipse around $\phi=0.5$ (for details on the various contributions see, e.g., \citealt{2012ApJ...761...53B}, \citealt{2010ApJ...725L.200S} or \citealt{2012ApJ...745...55G}): \smallskip

  (i) $F_e$, the ellipsoidal variations  resulting from tides on the star (of mass $M_{*}$ and radius $R_{*}$) raised by the planet of mass $M_p$ and semi-major axis $a$ described by:
\begin{eqnarray}
 F_{e} &=& -A_{e}[cos(2\pi 2\phi)+f_{1}cos(2\pi\phi)+f_{2}cos(2\pi 3\phi)]
\end{eqnarray}
where $f_{1}$ and $f_{2}$ are:
\begin{eqnarray}
f_{1}=3\alpha(a/R_{*})^{-1} \frac{5 sin^2(i)-4}{sin(i)}
\end{eqnarray}
\begin{eqnarray}
f_{2}=5\alpha(M_{p}/M_{*})(a/R_{*})^{-3} sin(i)
\end{eqnarray}
The parameter $\alpha$ is defined as
\begin{eqnarray}
\alpha=\frac{25u}{24(15+u)}\frac{y+2}{y+1}
\end{eqnarray}
where \textit{u} is the linear limb-darkening parameter and \textit{y} is the gravity darkening parameter; \smallskip

(ii) $F_d$, the Doppler boosting with amplitude $A_d$ caused by the host star's changing radial velocity described by:
\begin{eqnarray}
  F_{d} &=& A_d sin (2 \pi \phi)
\end{eqnarray}

\smallskip

(iii) $F_p$, the planet's phase function modeled as the variation in reflected light from a Lambertian sphere
(\citealt{1916ApJ....43..173R}) described by:

\begin{eqnarray}
  F_{p} &=& A_p \frac{sin (z) + (\pi - z) cos(z)}{\pi}
\end{eqnarray}

Here $A_p$ is the amplitude of the planetary phase function, and $z$ is related to phase $\phi$ and inclination $i$ via:

\begin{eqnarray}
 cos (z) &=& -sin(i)cos(2\pi\phi)
\end{eqnarray}\smallskip

(iv) $F_{ecl}$, the secondary eclipse -- i.e., the light that is blocked during the planet's passage behind its host star -- modeled using the description in \cite{2013ApJ...767...64R}, with $r(\phi)$, the separation between star and planet disk centers in the plane of the sky, in units of $R_{*}$, which we obtained from the referenced literature:

\begin{eqnarray}
 r(\phi)=(a/R_{*})[1-sin^2(i)cos^2(\phi-\phi_{m})]^{1/2}
\end{eqnarray}
where $\phi_{m}$ is the phase of the mid-point of the secondary eclipse.

$P_{ecl}$ is the eclipsed portion of the planet: 

\begin{eqnarray}
P_{ecl}(r)=\left\{ \begin{array}{lr}
	0 & :r\ge 1+p\\
	f(\theta_1,\theta_2)&: 1-p < r < 1+p\\
	1 & : r \le 1-p \end{array} \right.
\end{eqnarray}

where p is $\dfrac{R_p}{R_{*}}$ and with

\begin{eqnarray}
f(\theta_1,\theta_2) & =\frac{1}{\pi p^2} (\theta_1 -sin\theta_1 cos\theta_1)\\ &+\frac{1}{\pi}(\theta_2 - sin\theta_2 cos\theta_2)
\end{eqnarray}

Here  $\theta_{1}$ and $\theta_{2}$ are defined as:

\begin{eqnarray}
cos\theta_1=\dfrac{1+r^2-p^2}{2r}  &; 
  cos\theta_2=\dfrac{r^2+p^2-1}{2rp}
\end{eqnarray}

Hence the contribution $F_{ecl}$ of the secondary eclipse with depth $D_{ecl}$ is:
  
\begin{eqnarray}
F_{ecl}(\phi)=D_{ecl}[1-P_{ecl}(\phi, \phi_m)]
\end{eqnarray}

Here $D_{ecl}$ is a positive value, with $F_{ecl} = D_{ecl}$ outside of secondary eclipse, such that the planet's light is visible for most of the orbit, and then $F_{ecl}l = 0$ during secondary eclipse, as the planet's light is no longer visible.

\subsection{Lightcurve fitting}
We fit the cleaned (detrended and normalized with \texttt{kepflatten}) and phase-folded light-curves using MPFIT, an IDL package that implements Levenberg-Marquardt non-linear least squares curve fitting.

The errors on the parameters were obtained via the Levenberg-Marquardt fitting procedure, i.e., via the covariance matrix, which in some cases are likely to be underestimated as this method does not take into account parameter correlations.
 The period was held constant because we were fitting phase-folded data. The limb and gravity darkening parameters were trilinearly interpolated from the tables of \cite{2011AandA...529A..75C} and held constant during the fitting. The values for size ratio $R_p/R_*$  and the distance between the planet and the star $a/R_*$  as well as the used stellar parameters (e.g., $[Fe/H]$ or 
$log(g)$) were obtained from the NASA Exoplanet Archive or the literature (see Table \ref{tbl:used_params}) and fixed in the fitting procedure. Given the short periods and tidal circularization timescales, we assumed a circular orbit for higher order effects (e.g., of $e$ or $\omega$) in the contributions of the various phase curve effects, so that they keep their analytic form. The contribution of these second-order effects is negligible and no error leakage occurred (with the exception of KOI-13, which has a significant eccentricity, see Figure \ref{fig:fit1_2_10_13} and Section \ref{sec:koi-13}). The depth of the secondary eclipse was constrained to be positive.   Other parameters in the fit were the amplitude of the phase curve, the amplitude of the Doppler boosting, the amplitude of the ellipsoidal variations, the inclination, and the phase of the secondary eclipse. Restricting the depth of the secondary eclipse to be positive, along with the other fitted amplitudes, usually imparts a modeling bias towards higher positive values and more significant detections, especially if the signal is just above the noise level. Therefore we were very critical with marginal detections and tended towards null detections if the detected number was not significantly above the error level (see also discussion below and tables \ref{tab:lumresults1} and \ref{fig:non_detect}). For example in certain cases the fitted values for some of the phase contributions converged in the (non-zero) minimum value of $10^{-7}$ -- in other words these contributions were not needed to fit the data. In these cases we inserted a conservative $< 1 \ ppm$ in table \ref{tab:lumresults1}.

 In order to work on a uniform dataset and due to the computational intensity of fitting unbinned data, we cut out the transit part and binned all folded lightcurves down to 400 points for phase $\phi$=[0.1,0.9] (the \texttt{kepfold} routine used in the previous step already ensured that short- and long-cadence data were combined properly in an error-weighted way). While \citep{2010MNRAS.408.1758K} showed that binning can induce morphological
distortions to the photometric data light curve data of long cadence, we argue that the phase curve models work without a correction for this, as phase curve durations are significantly greater than the 30 minute cadence. For the secondary eclipse depths the integration time and binning will extend the secondary eclipse ingress and egress time, but those are very short compared to the eclipse duration.

We sought to include the best available parameters for the host stars in our modeling. Values from the Kepler Input Catalog (KIC) were often inaccurate, so we relied on the stellar parameters derived in the (mostly spectroscopic) planet confirmation observations (see Table\ref{tbl:used_params}) available at the time of our analysis.

\subsection{Temperatures and albedos}

We calculated the brightness temperatures using:
\begin{eqnarray}
D_{ecl}=(R_{p}/R_{*})^2 \frac{\int B_{\lambda}(T_{b})T_{K}d\lambda}{\int B_{\lambda}(T_{*})T_{K}d\lambda}
\end{eqnarray}
  where $D_{ecl}$ is the depth of the secondary eclipse, $R_{p}/R_{*}$ is the size ratio, $B_{\lambda}$ is the Planck function, $T_{b}$ is the brightness temperature, $T_{K}$ is the \textsl{Kepler} response function, and $T_{\star}$ is the stellar temperature. To solve for $T_{b}$, we integrated the right-hand side of
\begin{eqnarray}
\int B_{\lambda}(T_{b})T_{K}d\lambda=D_{ecl}(R_{*}/R_{p})^2\int B_{\lambda}(T_{*})T_{K}d\lambda
\end{eqnarray}
and then numerically integrated  the left-hand side iteratively using successively larger temperature values, until we found the brightness temperature that best matched the measured depth of the secondary eclipse. The nightside temperatures are calculated in the same way, but using $F_{night} = F_{ecl} - A_{p}$ instead.  If the difference between the phase curve and secondary eclipse was zero or negative, we did not calculate a night side temperature. When the derived error bounds for some of the candidates included zero we did not give a lower limit for the night side temperature.

Furthermore, we calculated the geometric albedo using:

\begin{eqnarray}
D_{ecl}=A_{g,obs}(R_{p}/a)^2
\end{eqnarray}
 which assumes no contribution from
thermal emission.

Correcting for thermal emission we used (see \citealt{2013ApJ...777..100H}):

\begin{eqnarray}
A_{g,corr}=A_{g,obs}-\frac{\pi \int B_{\lambda} (T_{eq}) d\lambda}{F_{0}} (a/R_{*})^2
\end{eqnarray}
where
\begin{eqnarray}
F_{0}=\sigma T_{0}^4
\end{eqnarray}
with
\begin{eqnarray}
T_{0}=T_{*} (R_{*}/a)^{1/2}
\end{eqnarray}

 Assuming a Lambertian criterion [$A_b=(3/2)A_g$] we calculated the equilibrium temperature using
\begin{eqnarray}
T_{eq}=T_{*}(f_{dist}R_{*}/a)^{1/2}(1-A_{b})^{1/4}
\end{eqnarray}

The resulting albedos corrected for thermal emission for no redistribution, $f_{dist}=\frac{1}{2}$, and fully efficient redistribution, $f_{dist}=\frac{2}{3}$, of heat from planetary day to night side are shown in Table \ref{tab:corralb}.

Cases where our derived night side temperatures resulted in upper limits indicate that the error in the night side flux was as large as the night side flux itself.

\subsection{Phase shift due to clouds}

For three of our targets (Kepler 7b, 12b and 43b) the fits using the phase curve model above were not sufficient and a clear systematic offset was seen in the residuals.
A quick literature search for these targets from our sample showed that this effect can be explained by a shift of the reflection signal, due to the superrotation phenomenon \citep{2013ApJ...776L..25D, 2013ApJ...777..100H}. In order to also model this higher order effect for these few exceptional systems, we added a component to the model lightcurve for (only) these three targets. This shift in the phase curve caused by planetary clouds (\citealt{2013ApJ...776L..25D}) for Kepler 7b, 12b and  43b was modeled in a similar fashion to the phase curve itself, where we added an additional free parameter $\phi_{shift}$ to describe the phase shift:

\begin{eqnarray}
 cos (z) &=& -sin(i)cos(2\pi(\phi+\phi_{shift}))
\end{eqnarray}\smallskip

Using this model to modify $F_p$ (updating equation 8 with equation 22) we found significant phase shifts $\phi_{shift}$ for KOI-20.01,  KOI-97.01 and KOI-135.01 (see \ref{sec:koi20}, \ref{sec:koi97}, and \ref{sec:koi135}).

\subsection{Upper limits}

For some of the planets in our sample we did not detect a secondary eclipse and/or a phase curve. In these cases we were only able to give upper limits for the derived parameters (see Section \ref{sec:non_detect}). In other cases our fits only constrained a limited number of parameters. For the planets in our sample for which the model does not fit a significant secondary eclipse, the upper limits for the secondary eclipse depths were calculated by adding the 1 sigma errors from the co-variance matrix to the best fit values. The same was applied and propagated  for the then derived albedo and brightness temperature limits of these candidates (see Section \ref{sec:non_detect}).

\section{Results}

Our results are summarized in Tables \ref{tab:lumresults1} to \ref{tab:uplim}. Figures \ref{fig:fit1_2_10_13} to \ref{fig:non_detect} illustrate our fitting efforts. In the following subsections we report individually on all our analyzed targets and, where possible, compare to previous observations and measurements.

\subsection{Detected secondary eclipses}

 \subsubsection{KOI 1.01 / TrES-2b / Kepler 1b}

TrES-2b (or KOI 1.01, Kepler-1b) -- a 1.28 $M_{Jup}$ and 1.24 $R_{Jup}$ planet on a 2.47 day orbit around a G0V star \citep{2006ApJ...651L..61O} -- was
 the first planet detected in the \textsl{Kepler} field.
    \cite{2011MNRAS.417L..88K} determine a day night contrast amplitude of 6.5   $\pm$   1.9 ppm which corresponds to a geometric albedo of $A_g$ = 0.025  $\pm$   0.007 and found a non-significant eclipse with depth of 16   $\pm$   13 ppm, similar to  \cite{2011ApJ...730...50K}, who derived a value of 21   $\pm$   22.  \cite{2011IAUS..276..475D} found  a geometric albedo of 0.06   $\pm$   0.05 and an equilibrium temperature of 1464 K using only Q1 data. \cite{2012ApJ...761...53B} derived a phase curve with  ellipsoidal variations and Doppler beaming of amplitudes  $2.79^{+0.44}_{-0.62}$ and $3.44^{+0.32}_{-0.37}$ ppm), respectively, and a difference between the day and night side planetary flux of $3.41^{+0.55}_{-0.82}$ ppm. They found a geometric albedo of $0.013^{+0.002}_{-0.003}$ and a secondary eclipse depth of $6.5^{+1.7}_{-1.8}$ ppm.   \cite{2012ApJ...761...53B} also showed that an atmosphere model that contains a temperature inversion is strongly preferred and  suggested that the \textsl{Kepler} bandpass probes a significantly greater atmospheric depth on the night side. The analysis of  \cite{2013ApJ...772...51E}
for TrES-2b shows an eclipse depth of 7.5   $\pm$   1.7 ppm,
a brightness temperature $T_b$ of $1910^{+40}_{-50}$K
a very low geometric albedo $A_g$ of 0.03  $\pm$  0.001
and a night side temperature
$T_{night}$ of 1700 K. For KOI-1.01 \citealt{2012AJ....143...39C} \footnote{When we cite values from \citealt{2012AJ....143...39C} we use the results for their method 8 - \textit{"CLM Light Curve with Eccentricity Free and Stellar Parameters 
        from from Isochrones"} - from the supplement materials of their paper.} measure an eclipse depth of $-9.3\pm 14.2$ ppm and report a maximum albedo of $-0.010^{+0.04}_{-0.040}$ and brightness temperature of $  -1700^{+3563}_{-243}$ K.

  Our fit for KOI-1.01 shows an eclipse depth of 10.9   $\pm$    2.3 ppm, a brightness temperature $T_b$ of $1901^{+27}_{-31}$  a geometric albedo $A_g$ of 0.05   $\pm$  0.01 a bond albedo $A_b$ of 0.08   $\pm$    0.02, leading to an upper limit for the night side temperature $T_{night}$ of $1885^{+ 51}_{- 66}$ K. These results agree very well with and therefore confirm the aforementioned measurements. Figure \ref{fig:fit1_2_10_13} (top, left) shows our fit results for KOI-1.01.

 \subsubsection{KOI 2.01 / HAT-P-7b / Kepler 2b}
  HAT-P-7b is a 1.78 $M_{Jup}$ and 1.36 $R_{Jup}$ planet on a 2.204 orbit around an evolved F6 star \citep{2008ApJ...680.1450P}. Due to its bright host star and its detection before the launch of the \textsl{Kepler} mission it is one of the best studied planets in our sample. A number of other groups already analyzed the secondary eclipse of this target using different methods with results spanning from 67 up to 130 ppm. From the first 10 days of \textsl{Kepler} calibration data \cite{2009Sci...325..709B} derived an eclipse depth of 130   $\pm$   11 ppm percent in the \textsl{Kepler} band. Using the whole first quartile Q1 data, \cite{2011IAUS..276..475D} derived a geometric albedo of 0.20   $\pm$   0.1 and an equilibrium temperature of 2085 K . \cite{2013ApJ...772...51E} measure an eclipse depth of 68.31   $\pm$  0.69 ppm, a brightness temperature $T_b$ of $2846^{+4}_{-4}$K a geometric albedo $A_g$ of 0.196  $\pm$  0.002 and an upper limit for the night side temperature $T_{night}$ of 1950 K for HAT-P-7. For KOI-2.01 \citealt{2012AJ....143...39C} measure an eclipse depth of $ 57.7\pm 9.30$ ppm and report a maximum albedo of $  0.37^{+0.07}_{-0.060}$ and brightness temperature of $   2860^{+65.00}_{-66}$ K.

\begin{figure*}
  \centering
      \includegraphics*[width=\textwidth]{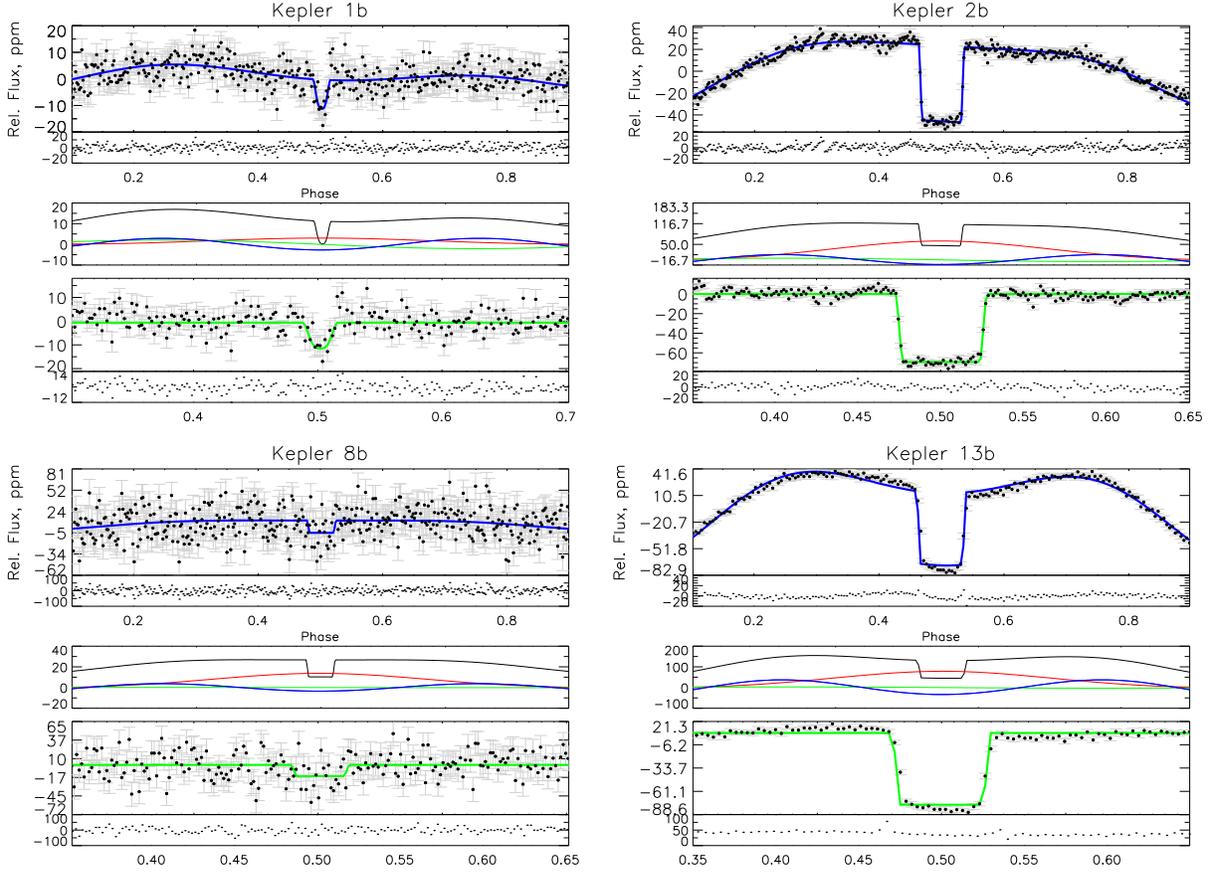}
       \caption{Fitted lightcurves for KOI-1.01/TrES-2b/Kepler 1b (top, left)  KOI-2.01/HAT-P-7b/Kepler 2b (top, right), KOI-10.01/Kepler 8b (bottom, left) and KOI-13.01/Kepler 13b (bottom, right). In each quarter: Phase curve, residuals (top).  Center: phase curve contributions: Doppler (blue), ellipsoidal (green), planetary phase (red). Bottom: zoom into phase curve subtracted secondary eclipse, residuals. Our model for KOI-13b shows a systematic offset from the actual data that may be explained by the system's eccentricity \citep{2014ApJ...795..112P} and/or gravity darkening \citep{2015ApJ...805...28M}.}
     \label{fig:fit1_2_10_13}
\end{figure*}

For KOI-2.01, we find an eclipse depth of 69.3   $\pm$    0.6 ppm, corresponding to a brightness temperature $T_b$ of $        2897^{+3}_{-4}$  K, a geometric albedo $A_g$ of 0.27   $\pm$     0.01 and a bond albedo $A_b$ of 0.4   $\pm$   0.01. The resulting upper limit for the night side temperature $T_{night}$ is  $2235^{+3 }_{-24 }$ K. Here - again - we are mostly in agreement with the other analyses from the literature. Figure \ref{fig:fit1_2_10_13} (top, right) shows our fit results for KOI-2.01.

  \subsubsection{KOI 10.01 / Kepler 8b}

       Kepler 8b, with a radius of 1.41 $R_{Jup}$ and a mass of 0.60 $M_{Jup}$, is among the lowest density planets (0.26 $\frac{g}{cm^3}$) known.  It orbits a relatively faint (V = 13.89 mag) F8IV subgiant host star with a period of P = 3.523 d and a semimajor axis of $0.0483^{+06}_{-0.0012}$ AU \citep{2010ApJ...724.1108J}. \cite{2011ApJ...730...50K}  exclude secondary eclipses of
depth 101.5 ppm or greater to 3-$\sigma$ confidence, which excludes a geometric albedo $>$ 0.63 to the same level. \cite{2011IAUS..276..475D} report a geometric albedo of 0.21   $\pm$   0.1 and an equilibrium temperature of 1567 K using only Q1 data. For Kepler 8b, \cite{2013ApJ...772...51E} derive an eclipse depth of 26.2   $\pm$  5.6 ppm, a brightness temperature $T_b$ of $2370^{+50}_{-70}$K a geometric albedo $A_g$ of 0.134  $\pm$  0.03 and an upper limit for the night side temperature $T_{night}$ of 2100 K. For KOI-10.01 \citealt{2012AJ....143...39C} measure an eclipse depth of $-5.8\pm 45.9$ ppm and report a maximum albedo of $  0.36^{+ 0.36}_{-0.33}$ and brightness temperature of $   2463^{+215.0}_{-587}$ K.

 In our fits of KOI-10.01 we find an eclipse depth of  16.5  $\pm$    4.45   ppm, a brightness temperature $T_b$ of $2241^{+61}_{-       77}$  K, a geometric albedo $A_g$ of       0.11   $\pm$      0.03, a bond albedo $A_b$ of 0.16   $\pm$      0.04 and an upper limit for the night side temperature $T_{night}$ of $1859^{+227 }$ K. Our results are slightly lower, however consistent with \citealt{2013ApJ...772...51E}. Figure \ref{fig:fit1_2_10_13} (bottom, left) shows our fit results for Kepler 8b.

 \subsubsection{KOI 13.01 / Kepler 13b}\label{sec:koi-13}
Kepler 13 (or KOI 13.01) is the second brightest host star in our sample with mag 9.958 in the \textsl{Kepler} band. The planet Kepler 13b was detected by \cite{2011AJ....142..195S} due to its photometric orbit using the BEER algorithm \citep{2011MNRAS.415.3921F}. KOI-13.01 is a super-Jovian planet with a mass of 8.3 $M_{Jup}$ and 1.4 $R_{Jup}$ radius on a 1.76 day orbit around a A5-7V host star \citep{2012MNRAS.422.1512M}. \cite{2012AandA...544L..12S} found that the transiting planet is orbiting the main component of a hierarchical triple system of two fast rotating a stars and one more companion with mass between 0.4 and 1 $M_{Sun}$. \cite{2012MNRAS.421L.122S} reported a spin-orbit resonance, transit duration variation and possible secular perturbations in the KOI-13 system. For KOI-13.01, \citealt{2013ApJ...772...51E} measure
an eclipse depth of 143.0   $\pm$  1.2 ppm,
a brightness temperature $T_b$ of $3706^{+5}_{-6}$K
a geometric albedo $A_g$ of 0.42   $\pm$  0.0031
and an upper limit for the night side temperature
$T_{night}$ of 2710. For KOI-13.01 \citealt{2012AJ....143...39C} measure an eclipse depth of $ 88.8\pm 6.19$ ppm and report a maximum albedo of $  0.58^{+0.03}_{-0.030}$ and brightness temperature of $   3758^{+55.00}_{-57}$ K.

For Kepler 13b, we find an eclipse depth of 84.8  $\pm$  5.4, corresponding to a brightness temperature of $       3421^{+          32}_{-          35}$, a geometric albedo of  0.27  $\pm$      0.02, a bond albedo of       0.40    $\pm$  
    0.03   and a night side temperature of  $2394^{+251}$. Figure \ref{fig:fit1_2_10_13} (bottom, right) shows our fit results for Kepler 13b.
    
    Part of the reason for the inconsistency with the \citealt{2013ApJ...772...51E} numbers is the different way the dilution between the two stars in the visual binary was accounted for also the slightly different host star parameters assumed in each paper. Our results compare very well to the numbers of \citealt{2014ApJ...788...92S}. However our fits show a clearly visual offset from the actual data (see \ref{fig:fit1_2_10_13}; bottom, right -- slope of the bottom of the secondary eclipse, excess flux prior to secondary). We suspect that the reported low but significant eccentricity of  $0.034 \pm 0.003$ \citep{2014ApJ...795..112P,2014ApJ...788...92S} and gravity darkening \citep[as shown in][]{2015ApJ...805...28M} can explain this effect.

 \subsubsection{KOI 17.01 / Kepler 6b}
  Kepler 6b \citep{2010ApJ...713L.136D} is a transiting Hot Jupiter orbiting a 3.8 Gyr old star with unusually high metallicity  ([Fe/H] = +0.34   $\pm$   0.04) in a  P = 3.235 d orbit. It has a mass of  $M_P$ = 0.67 $M_{Jup}$, and a radius of $R_P$ = 1.32 $R_{Jup}$, resulting in a density of 0.35 $(g/cm^3)$. The host star Kepler 6 is more massive and larger than the sun (1.21 $M_{Sun}$, 1.39 $R_{Sun}$) but slightly cooler with $T_s$=5724 K. \cite{2011ApJ...730...50K}  exclude secondary eclipses of
depth 51.5 ppm or greater to 3-$\sigma$ confidence, which excludes a geometric albedo of more than 0.32 to the same level. \cite{2011ApJS..197...11D} found a brightness temperatures of Kepler 6b from Spitzer observations of $T_B$ = 1660   $\pm$   120 K and an optical geometric albedo $A_g$ in the \textsl{Kepler} bandpass of $A_g$ = 0.11   $\pm$   0.04. \cite{2011IAUS..276..475D} derive a geometric albedo of 0.18   $\pm$   0.09 and an equilibrium temperature of 1411 K from \textsl{Kepler}'s Q1 data. For Kepler 6, \cite{2013ApJ...772...51E} measure an eclipse depth of 8.9   $\pm$  3.8 ppm,
a brightness temperature $T_b$ of $2000^{+80}_{-100}$K, a geometric albedo $A_g$ of 0.058   $\pm$  0.025 and an upper limit for the night side temperature $T_{night}$ of 1600 K.
For KOI-17.01 \citealt{2012AJ....143...39C} measure an eclipse depth of $-71.7\pm 33.6$ ppm and report a maximum albedo of $ -0.33^{+0.27}_{-0.27}$ and brightness temperature of $  -2306^{+369.0}_{-163}$ K.

\begin{figure*}
  \centering
      \includegraphics*[width=\textwidth]{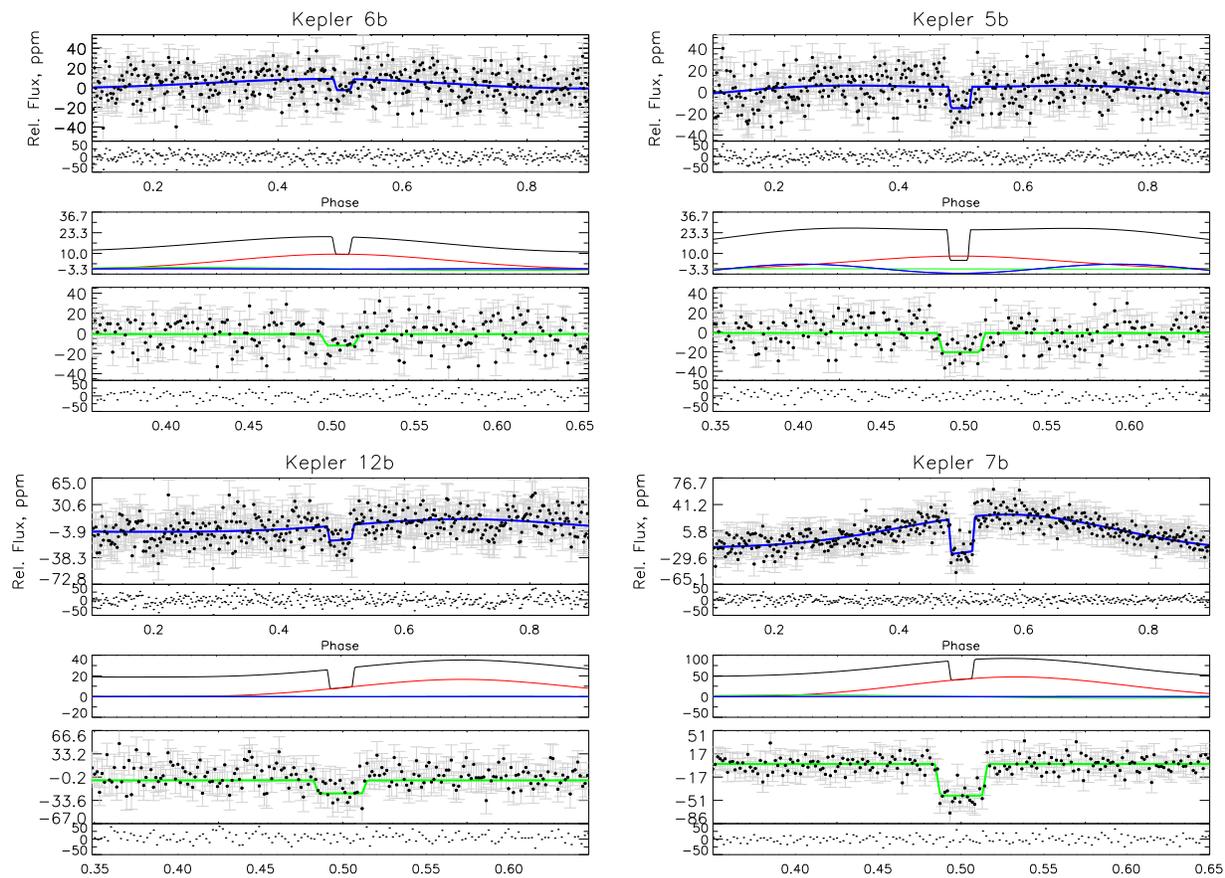}
       \caption{Fitted lightcurves for KOI-17.01/Kepler 6b (top, left)  KOI-18.01/Kepler 5b (top, right), KOI-20.01/Kepler 12b (bottom, left) and KOI-97.01/Kepler 7b (bottom, right). In each quarter: Phase curve, residuals (top).  Center: phasecurve contributions: Doppler (blue), ellipsoidal (green), planetary phase (red). Bottom: zoom into phase curve subtracted secondary eclipse, residuals.}
     \label{fig:fit17_18_20_97}
\end{figure*}

 For KOI-17.01, we derive an eclipse depth of        11.3   $\pm$   4.2 ppm, a brightness temperature $T_b$ of $2060^{+70}_{-95}$  K, a geometric albedo $A_g$ of 0.07  $\pm$   0.03, a bond albedo $A_b$ of 0.11   $\pm$   0.04 and an upper limit for the night side temperature $T_{night}$ of $1719^{+236}$ K - again confirming the results of \cite{2013ApJ...772...51E}. Figure \ref{fig:fit17_18_20_97} (top, left) shows our fit results for Kepler 6b.
 
     \subsubsection{KOI 18.01 / Kepler 5b}

     Kepler 5b (or KOI 18.01) is a 2.11 $M_{Jup}$ and 1.43 $R_{Jup}$ planet on a 3.55 day orbit around a 13th magnitude star \citep{2010ApJ...713L.131K}. \cite{2011ApJ...730...50K} detect a weak secondary eclipse for Kepler 5b of depth 25   $\pm$   17 ppm and a geometric albedo of $A_g$ = 0.15   $\pm$   0.10. \cite{2011ApJS..197...11D} found a brightness temperatures of Kepler 5b from Spitzer observations of $T_B$ = 1930   $\pm$   100 K and an optical geometric albedo  in the \textsl{Kepler} band of $A_g$ = 0.12   $\pm$   0.04. \cite{2011IAUS..276..475D} report a geometric albedo of 0.21   $\pm$   0.1 and an equilibrium temperature of 1557 K using only Q1 data. For Kepler 5b, \cite{2013ApJ...772...51E} find
an eclipse depth of 18.8   $\pm$  3.7 ppm,
a brightness temperature $T_b$ of $2400^{+50}_{-60}$K,
a geometric albedo $A_g$ of 0.119   $\pm$  0.025
and an upper limit for the night side temperature
$T_{night}$ of 2100. For KOI-18.01 \citealt{2012AJ....143...39C} measure an eclipse depth of $-89.8\pm 48.6$ ppm and report a maximum albedo of $ -0.47^{+0.51}_{-0.60}$ and brightness temperature of $  -2399^{+4274}_{-257}$ K.

Our lightcurve fits of KOI-18.01  show an eclipse depth of  19.8   $\pm$   3.65  ppm, a brightness temperature $T_b$ of $        2305^{+46}_{-52}$  K, a geometric albedo $A_g$ of 0.16   $\pm$   0.03, a bond albedo $A_b$ of 0.25   $\pm$   0.05 and a night side temperature $T_{night}$ of $2169^{+ 81}_{- 113}$ K. These results are very close to the values of \cite{2013ApJ...772...51E}. Figure \ref{fig:fit17_18_20_97} (top, right) shows our fit results for KOI-18.01.

      \subsubsection{KOI 20.01  / Kepler 12b}\label{sec:koi20}
          Kepler 12b (KOI 20.01, \citealt{2011ApJS..197....9F}), with a radius of 1.69   $\pm$   0.03 $R_{Jup}$ and a mass of 0.43  $\pm$  0.04 $M_{Jup}$, belongs to the group of planets with highly inflated radii. On a 4.44 day orbit around a slightly evolved G0 host, Kepler 12b is the least irradiated within the class of inflated and very low density planets  (0.11  $\pm$  0.01 $(g/cm^3)$) and may have important implications for the question of the correlation between irradiation and inflation.   \cite{2011ApJS..197....9F} also detected a secondary eclipse depth pf 31   $\pm$   7 ppm, corresponding to a geometric albedo of 0.14  $\pm$  0.04. For KOI-20.01 \citealt{2012AJ....143...39C} measure an eclipse depth of $-26.3\pm 28.5$ ppm and report a maximum albedo of $  0.16^{+ 0.17}_{-0.13}$ and brightness temperature of $   2115^{+180.0}_{-312}$ K.

 Our fits of KOI 20.01's lightcurve confirm and improve this with a resulting eclipse depth of  18.7   $\pm$   4.9 ppm, a brightness temperature $T_b$ of $      2121^{+54}_{-67}$  K, a geometric albedo $A_g$ of 0.09   $\pm$  0.02, a bond albedo $A_b$ of 0.14   $\pm$  0.04  and an upper limit for the night side temperature $T_{night}$ of $1711^{+ 223}$ K. Furthermore we found a phase shifts $\phi_{shift}$ of -0.19 $\pm$ 0.03 for KOI 20.01. Figure \ref{fig:fit17_18_20_97} (bottom, left) shows our fit results for Kepler 12b.

   \subsubsection{KOI 97.01 / Kepler 7b}\label{sec:koi97}

Kepler 7b (\citealt{2010ApJ...713L.140L}) with a mass  of  0.43 $M_{Jup}$ and radius 1.48 $R_{Jup}$ also has a very low density of $ 0.17 (g/cm^3)$. For KOI-97.01 \citealt{2012AJ....143...39C} measure an eclipse depth of $ 79.1\pm 15.3$ ppm and report a maximum albedo of $   1.26^{+0.36}_{-0.33}$ and brightness temperature of $   2626^{+90.00}_{-101}$. \cite{2011ApJ...735L..12D}, using Q0-Q4 data, measure an occultation depth in the \textsl{Kepler} bandpass of 44   $\pm$   5 ppm, a geometric albedo of 0.32   $\pm$   0.03, and a planetary orbital phase light curve with an amplitude of 42   $\pm$   4 ppm. 

Our results for KOI-97.01 confirm this almost exactly: we find an eclipse depth of        46.6   $\pm$   4.0  ppm, a brightness temperature $T_b$ of $2547^{+26}_{-         28}$  K, a geometric albedo $A_g$ of       0.32   $\pm$  0.03 and a bond albedo $A_b$ of 0.48   $\pm$  0.04. We also detected an additionally phase shifts $\phi_{shift}$ of  -0.08 $\pm$ 0.02 for KOI-97.01. Figure \ref{fig:fit17_18_20_97} (bottom, left) shows our fit results for Kepler 12b.
It is interesting to see that the error values for \cite{2011ApJ...735L..12D} are almost exactly the same, even though they only used Q0-4 in comparison to Q0-15 in our study. One possible explanation could be that \cite{2011ApJ...735L..12D} did not include Doppler or ellipsoidal components in their model or maybe these result hint at a systematic noise floor we are reaching here. 

We also confirm the following results: \cite{2011ApJ...730...50K} detect a secondary eclipse for Kepler 7b of depth 47   $\pm$   14 ppm and a geometric albedo of $A_g$ = 0.38   $\pm$   0.12.  The day-night difference of 17   $\pm$   9 ppm they calculate supports the hypothesis of thermal emission as a source for both the secondary eclipse and the phase curve. \cite{2011IAUS..276..475D} find a geometric albedo of 0.35   $\pm$   0.11 and an equilibrium temperature of 1370 K using only data from the first quartile.

      \subsubsection{KOI 127.01 / Kepler 77b}

Kepler 77b \citep{2013AandA...557A..74G} is a moderately bloated planet with a mass of $M_P$ = 0.430  $\pm$   0.032 $M_{Jup}$, a radius of $R_P$ = 0.960  $\pm$   0.016 $R_{Jup}$, orbiting) G5 V star with a period of 3.58 days. \cite{2013AandA...557A..74G} do not find a secondary eclipse with a depth larger than 10 ppm which leads to limits of the geometric and Bond albedo of $A_g$ $\leq$ 0.087  $\pm$   0.008 and $A_b$ $\leq$ 0.058  $\pm$   0.006, respectively. For KOI-127.01 \citealt{2012AJ....143...39C} measure an eclipse depth of $-92.6\pm 39.1$ ppm and report a maximum albedo of $  -1.11^{+0.36}_{-0.41}$ and brightness temperature of $  -2563^{+129.0}_{-112}$ K.

\begin{figure*}
  \centering
      \includegraphics*[width=\textwidth]{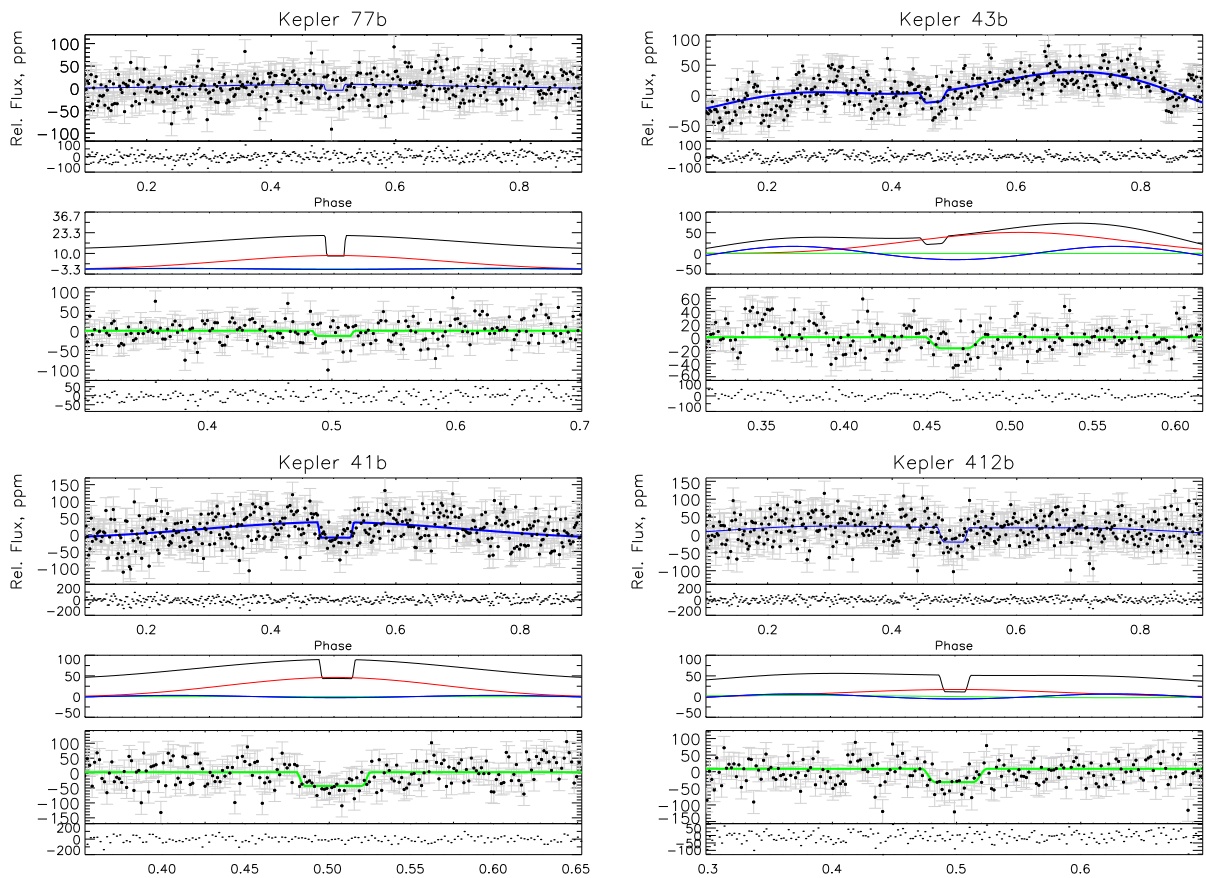}
       \caption{Fitted lightcurves for KOI-127.01/Kepler 77b (top, left),  KOI-135.01/Kepler 43b (top, right), KOI-196.01/Kepler 41b (bottom, left) and KOI-202.01/Kepler 412b (bottom, right). In each quarter: Phase curve, residuals (top).  Center: phasecurve contributions: Doppler (blue), ellipsoidal (green), planetary phase (red). Bottom: zoom into phase curve subtracted secondary eclipse, residuals.}
     \label{fig:fit127_135_196_202}
\end{figure*}

  For KOI-127.01, we find an eclipse depth of 13.3   $\pm$   7.4 ppm, which results in a brightness temperature of  $        2062^{+         100}_{-        165}$ K, geometric albedo of     0.15  $\pm$      0.09, bond albedo of      0.23  $\pm$  
     0.13   and night side temperature of  $1854^{+ 216}$ K.  Figure \ref{fig:fit127_135_196_202} (top, left) shows our fit results for Kepler 77b.

      \subsubsection{KOI 135.01 / Kepler 43b}\label{sec:koi135}

KOI-135.01 \citep{2012AandA...538A..96B} with radius $R_P$ = 1.20  $\pm$   0.06 $R_{Jup}$  and mass $M_P$ = 3.23  $\pm$   0.19 $M_{Jup}$ orbits its parent star in 3.02 days. We are the first to report a secondary eclipse of KOI-135.01 with a depth of 17.0   $\pm$   5.3  ppm. This corresponds to a brightness temperature $T_b$ of $        2296^{+73}_{-95}$ K, a very low geometric albedo $A_g$ of  0.06   $\pm$   0.02 and a bond albedo $A_b$ of  0.09   $\pm$      0.03 respectively. Using our model we also found a phase shift $\phi_{shift}$ of -0.10 $\pm$ 0.01 for KOI-135.01. Figure \ref{fig:fit127_135_196_202} (top, right) shows our fit results for KOI-135.01.

      \subsubsection{KOI 196.01 / Kepler 41b}

     The planet KOI-196.01, with a radius of 0.84  $\pm$  0.03 $R_{Jup}$ and a mass of 0.49  $\pm$  0.09 $M_{Jup}$, orbits a G2V star of 0.99  $\pm$  0.03 $R_{sun}$ (\citealt{2011AandA...536A..70S}): KOI-196.01 is one the rare close-in Hot Jupiters with a radius smaller than Jupiter suggesting a non-inflated planet. \cite{2011AandA...536A..70S} detect a secondary eclipse depth of 64   $\pm$   10 ppm as well as the optical phase variation, leading to a relatively high geometric albedo of $A_g$ = 0.3  $\pm$  0.08 and a temperature of $T_B$ = 193  $\pm$  80 K. \cite{2013ApJ...767..137Q}  confirmed the Hot Jupiter Kepler 41b via phase curve analysis and find a secondary eclipse depth of 60
  $\pm$   9 ppm and a geometric albedo of
$A_g$ = 0.23  $\pm$  0.05. For KOI-196.01 \citealt{2012AJ....143...39C} measure an eclipse depth of $ 75.4\pm 40.1$ ppm and report a maximum albedo of $  0.33^{+ 0.16}_{-0.15}$ and brightness temperature of $   2463^{+130.0}_{-168}$ K.

   For KOI-196.01, we find an eclipse depth of        46.2   $\pm$   8.7  ppm, slightly lower than, but however consistent with \cite{2013ApJ...767..137Q} and \cite{2011AandA...536A..70S}. Our fits correspond to a brightness temperature $T_b$ of $  2395^{+50}_{-58}$  K, a geometric albedo $A_g$ of  0.18   $\pm$   0.03 and a bond albedo $A_b$ of 0.27   $\pm$      0.05. Figure \ref{fig:fit127_135_196_202} (bottom, left) shows our fit results for KOI-196.01.

     \subsubsection{KOI 202.01 / Kepler 412b}

 The planet Kepler 412b \citep{2014arXiv1401.6811D} is an inflated Jupiter with a mass of 0.94   $\pm$   0.09 $M_{Jup}$ and a radius of 1.33   $\pm$   0.04 $R_{Jup}$ orbiting its G3 V host star in 1.72 days . \cite{2014arXiv1401.6811D} detected a secondary eclipse 47.4   $\pm$  
7.4 ppm and derived the day side temperature to be a maximum of 2380
  $\pm$   40 K and estimated the geometrical albedo, $A_g$, in the range 0.09 to 0.13 and a night side temperature of 2154   $\pm$   83 K. For KOI-202.01 \citealt{2012AJ....143...39C} measure an eclipse depth of $ 63.6\pm 37.7$ ppm and report a maximum albedo of $  0.19^{+ 0.1}_{-0.10}$ and brightness temperature of $   2455^{+138.0}_{-207}$ K.
   Figure \ref{fig:fit127_135_196_202} (bottom, right) shows our fit results for Kepler 412b. Our analysis of Kepler 412b shows an eclipse depth of 40.2   $\pm$   9.0 ppm, well within the lower end of the value of  \cite{2014arXiv1401.6811D}. We therefore get a brightness temperature of $        2355^{+          35}_{-          40}$ K, geometric albedo of   0.11  $\pm$      0.02 , bond albedo of  0.16    $\pm$  0.04  and a night side temperature of $2210^{+105 }_{- 163}$ K.

     \subsubsection{KOI 203.01 / Kepler 17b}

     Kepler 17b is a  $M_P$=2.45   $\pm$   0.11 $M_{Jup}$ and $R_P$=1.31   $\pm$  0.02 $R_{Jup}$ planet orbiting a 1.02  $\pm$  0.03 $R_{Sun}$ star with a period of 1.49 days \citep{2011ApJS..197...13E}.  \cite{2011ApJS..197...13E} find measure an eclipse depth of 58  $\pm$  10 ppm and a geometric albedo $A_g$ of 0.1  $\pm$  0.02. \cite{2012AandA...538A..96B} find a slightly different $M_p$ = 2.47   $\pm$   0.10 $M_{Jup}$ and $R_p$ = 1.33   $\pm$   0.04 $R_{Jup}$ and an upper limit for the geometric albedo of $A_g < 0.12 $. For KOI-203.01 \citealt{2012AJ....143...39C} measure an eclipse depth of $ 52.\pm 98.0$ ppm and report a maximum albedo of $  0.14^{+ 0.14}_{-0.15}$ and brightness temperature of $   2319^{+202.0}_{-4138}$ K.

\begin{figure*}
  \centering
      \includegraphics*[width=\textwidth]{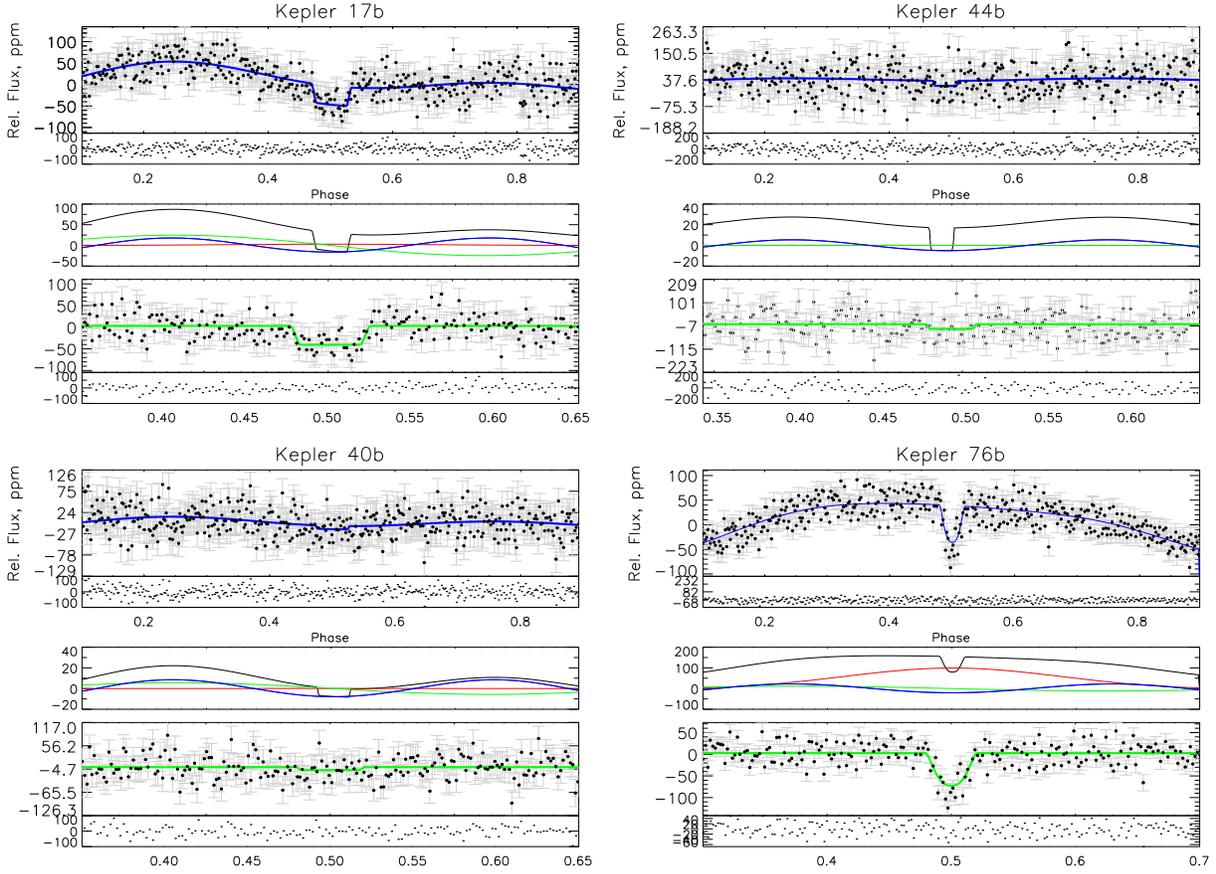}
       \caption{Fitted lightcurves for KOI-203.01/Kepler 17b (top, left)  KOI-204.01/Kepler 44b (top, right), KOI-428.01/Kepler 40b (bottom, left) and KOI-1658.01/Kepler 76b (bottom, right). In each quarter: Phase curve, residuals (top).  Center: phasecurve contributions: Doppler (blue), ellipsoidal (green), planetary phase (red). Bottom: zoom into phase curve subtracted secondary eclipse, residuals.}
     \label{fig:fit203_204_428_1658}
\end{figure*}

For KOI-203.01, we find an eclipse depth of    43.7   $\pm$   6.4  ppm, a brightness temperature $T_b$ of $ 2247^{+35}_{-40}$  K, a geometric albedo $A_g$ of 0.08   $\pm$   0.01 a bond albedo $A_b$ of 0.13   $\pm$   0.02 and an upper limit for the night side temperature $T_{night}$ of $2229^{+50 }_{-58 }$ K. These results are slightly different from, but still consistent with \cite{2012AandA...538A..96B} and \cite{2011ApJS..197...13E}. Figure \ref{fig:fit203_204_428_1658} (top, left) shows our fit results for KOI 203.01.

      \subsubsection{KOI 204.01 / Kepler 44b}

       KOI-204.01  \citep{2012AandA...538A..96B} is a 1.24  $\pm$   0.07 $R_{Jup}$, 1.02  $\pm$   0.07 $M_{Jup}$ planet orbiting its parent G2IV star in   3.25 days. For KOI-204.01 \citealt{2012AJ....143...39C} measure an eclipse depth of $ 54.8\pm 82.5$ ppm and report a maximum albedo of $   1.92^{+ 1.670}_{- 1.74}$ and brightness temperature of $   2669^{+231.0}_{-641}$ K.

 For KOI-204.01, we marginally detect a secondary eclipse with a depth of  21.9   $\pm$         14.9  ppm. This corresponds to a brightness temperature $T_b$ of $ 2348^{+149}_{-         279}$  a geometric albedo $A_g$ of 0.28   $\pm$       0.19, a bond albedo $A_b$ of
 0.42   $\pm$  0.29 and an upper limit for the night side temperature $T_{night}$ of $2347^{+149 }_{-280 }$ K. Figure \ref{fig:fit203_204_428_1658} (top, right) shows our fit results for KOI 204.01.


%
%
%

  \subsubsection{KOI 428.01 / Kepler 40b}

 The planet KOI-428.01 (1.17   $\pm$   0.04 $R_{Jup}$; 2.2   $\pm$   0.4 $M_{Jup}$), orbits an F5IV star of 2.13   $\pm$   0.06 $R_{Sun}$, 1.48   $\pm$   0.06 $M_{Sun}$, one of the largest and the most evolved stars discovered so far with a transiting planet \citep{2011AandA...528A..63S}.

For KOI-428.01, we detect an eclipse depth of        7.91   $\pm$   7.55 ppm, consistent with a non detection within one sigma. This corresponds to limits for the brightness temperature $T_b$ of $        2331^{+193}_{-626}$  K, the geometric albedo $A_g$ of  0.09   $\pm$      0.08,  the bond albedo $A_b$ of 0.13   $\pm$       0.13 and an upper limit for the night side temperature $T_{night}$ of $2327^{+195 }_{-669 }$ K. Figure \ref{fig:fit203_204_428_1658} (bottom, left) shows our fit results for KOI 428.01.




%
%

 \subsubsection{KOI 1658.01 / Kepler 76b}

Kepler 76b \citep{2013ApJ...771...26F}
(2.0  $\pm$  0.26 $M_{Jup}$, 1.25   $\pm$   0.08 R$_{Jup}$) orbits a 1.2 $M_{Sun}$ star in 1.55 days.
 It is slightly denser than Jupiter indicating that it is not inflated like other planets in this sample. \cite{2013ApJ...771...26F} find a secondary eclipse depth of 98.9   $\pm$   7.1 ppm as well as significant contribution to doppler, elipsoidal and phase modulatios of 13.5, 21.1 and 50.4 ppm.

 For Kepler 76b, our model fits an eclipse of 75.6   $\pm$   5.6 ppm, about 25$\%$ less than in   \cite{2013ApJ...771...26F}. Using our values we find a brightness temperature of $        2776^{+         26}_{-         28}$ K, a geometric albedo of    0.22  $\pm$      0.02  and bond albedo of 0.33    $\pm$   0.02. Figure \ref{fig:fit203_204_428_1658} (bottom, right) shows our fit results for KOI 1658.01. We find a Doppler boosting of  11.4 $\pm$ 1.0  and a ellipsoidal variation of 22.6 $\pm$ 1.9,       well in agreement with \citep{2013ApJ...771...26F}. However, our derived phasecurve amplitude of 101.3 $\pm$ 3.6 is a factor two bigger than theirs. Reasons for this discrepancy -- also in the derived eclipse depth -- could be third light contributions \citep[$f_3$ = 0.056, average third light fraction used in][]{2013ApJ...771...26F} or different stellar parameters \citep[we used their TODMOR derived values from table 2 in][whereas they fitted to the values in table 1]{2013ApJ...771...26F}.

\begin{deluxetable}{lcccc}
\tabletypesize{\footnotesize}
\tablecolumns{10}
\tablewidth{0pt}
\tablecaption{Fit results for the amplitudes of phasecurve ($A_p$), Doppler boosting ($A_d$) and ellipsoidal variation ($A_e$), and eclipse depth $D_{ecl}$ - all in ppm
\label{tab:lumresults1}}

\tablehead{\colhead{KOI} & \colhead{$A_p$}   & \colhead{$A_d$} &\colhead{$A_e$} &  \colhead{$D_{ecl}$}}

\startdata 

   1 &        3.0  $\pm$        0.8 &       2.0  $\pm$        0.2 &        2.9  $\pm$  
      0.5 &        10.9  $\pm$         2.2 \\
           2 &        60.8  $\pm$        0.5 &       5.3  $\pm$        0.1 &        16.8  $\pm$  
      0.3 &        69.3  $\pm$        0.5 \\
          10 &        13.8  $\pm$         3.7 &       $<1$  \tablenotemark{a} &        3.7  $\pm$  
       2.0 &        16.5  $\pm$         4.4 \\
        13 &        78.7  $\pm$        5.4 &       $<1$  \tablenotemark{a} &        37.1  $\pm$  
       5.4 &        84.8  $\pm$         5.4 \\
       
          17 &        9.5  $\pm$         2.7 &      1.0  $\pm$        0.9 &      $<1$  \tablenotemark{a} &        11.3  $\pm$         4.2 \\
          18 &        8.3  $\pm$         2.6 &       $<1$  \tablenotemark{a} &        3.1  $\pm$  
       1.4 &        19.8  $\pm$         3.6 \\
          20 &        16.7  $\pm$         2.6 &       $<1$  \tablenotemark{a} &       $<1$  \tablenotemark{a} &        18.7  $\pm$         4.9 \\
          97 &        47.8  $\pm$         5.2 &       3.9  $\pm$         4.0 &       $<1$  \tablenotemark{a} &        46.6  $\pm$         3.9 \\
         127 &     8.6  $\pm$         5.4 &     $<1$  \tablenotemark{a} &      $<1$  \tablenotemark{a} &        13.3  $\pm$         7.4 \\
         135 &        51.7  $\pm$         3.3 &       $<1$  \tablenotemark{a} &        16.9  $\pm$  
       2.2 &        17.0  $\pm$         5.3 \\
         196 &        46.3  $\pm$         7.9 &       $<1$  \tablenotemark{a} &        2.8  $\pm$  
       4.3 &        46.2  $\pm$         8.7 \\
              202 &        17.3  $\pm$         7.4 &       2.7  $\pm$         2.2 &        6.3  $\pm$   3.9 &        40.2  $\pm$         9.0 \\
         203 &        2.9  $\pm$         5.9 &       24.7  $\pm$         1.8 &        17.9  $\pm$   3.2 &        43.7  $\pm$         6.4 \\
         204 &        $<1$  \tablenotemark{a} &       $<1$  \tablenotemark{a} &        5.4  $\pm$  
       4.8 &        21.9  $\pm$         14.9 \\
         428 &        $<1$  \tablenotemark{a} &       5.6  $\pm$         2.1 &        8.5  $\pm$  
       2.9 &        7.9  $\pm$         7.5 \\
         1648 &      101.3   $\pm$   3.6  &      11.4   $\pm$      1.0 &        22.6   $\pm$  
       1.9 &        75.6  $\pm$         5.6 
\enddata
\vspace{0.02cm}

\tablenotetext{a}{Equivalent to not detectable in our fit. In these cases the fitted values of the phase contributions converged in the (non-zero) minimum value of $10^{-7}$ - in other words these contributions were not needed to fit the data. Therefore we put in a conservative limit of $< 1 ppm$.}
\end{deluxetable}

\begin{deluxetable}{lcccccc}
\tabletypesize{\footnotesize}
\tablecolumns{10}
\tablewidth{0pt}
\tablecaption{Derived brightness temperatures ($T_b$), geometric and bond albedos ($A_g$,$A_b$), equilibrium temperatures $T_{eq}$ for no $f_{dist}=\frac{1}{2}$ and full redistribution $f_{dist}=\frac{2}{3}$  and night side temperature $T_{night}. $
\label{tab:tempresults}}
\tablehead{\colhead{KOI} & \colhead{$T_b$}   & \colhead{$A_g$\tablenotemark{a}} &\colhead{$A_b$} &\colhead{$T^{eq}_{1/2}$}&\colhead{$T^{eq}_{2/3}$}&\colhead{$T_{night}\tablenotemark{b}$ [K]}
}

\startdata
 1 & $        1947^{+          37}_{-          45}$ &     0.05  $\pm$      0.01   &      0.08  $\pm$  
    0.02   &         1363 &         1574 &       $1885^{+ 51}_{- 66}$\\
           2 & $        2897^{+           3}_{-           3}$ &      0.27    $\pm$       0.003 &       0.4  $\pm$  
     0.003 &         1892 &         21858 &       $2235^{+3 }_{-24 }$ \\
          10 & $        2241^{+          61}_{-          76}$ &      0.11    $\pm$      0.03   &       0.16    $\pm$  
    0.04   &         1536 &         1774 &       $1859^{+227 }$ \\
     13 & $       3421^{+          32}_{-          35}$ &     0.27  $\pm$      0.02   &       0.40    $\pm$  
    0.03   & 2620 &         3025 &       $2394^{+251}$ \\
          17 & $        2060^{+          70}_{-          96}$ &     0.07  $\pm$      0.03   &       0.11    $\pm$  
    0.04   &         1413 &         1632 &       $1719^{+236}$ \\
          18 & $        2305^{+          45}_{-          53}$ &      0.16    $\pm$      0.03   &       0.25    $\pm$  
    0.05   &         1429 &         1650 &       $2169^{+ 81}_{- 113}$ \\
          20 & $        2120^{+          54}_{-          67}$ &     0.09  $\pm$      0.02   &       0.14    $\pm$  
    0.04   &         1422 &         1642 &       $1711^{+ 223}$ \\
          97 & $        2547^{+          26}_{-          28}$ &      0.32    $\pm$      0.03   &       0.48    $\pm$  
    0.04   &         1364 &         1575 &      -- \\
         127 & $        2062^{+         100}_{-        165}$ &     0.15  $\pm$      0.09   &      0.23  $\pm$  
     0.13   &         1136 &         1312 &       $1854^{+ 216}$ \\
         135 & $        2295^{+          74}_{-          95}$ &     0.06  $\pm$      0.02   &      0.09  $\pm$  
    0.03   &         1930 &         2229 &     -- \\
         196 & $        2395^{+          50}_{-          57}$ &      0.18    $\pm$      0.03   &       0.27    $\pm$  
    0.05   &         1513 &         1933 &     -- \\
     202 & $        2355^{+          35}_{-          40}$ &     0.11  $\pm$      0.02   &       0.16    $\pm$  
    0.04   &  1701 &         1965 &       $2210^{+105 }_{- 163}$ \\
         203 & $        2247^{+          35}_{-          40}$ &     0.08  $\pm$      0.01   &       0.13    $\pm$  
    0.02   &         1660 &         1917 &       $2229^{+50 }_{-58 }$ \\
         204 & $        2348^{+         149}_{-         277}$ &      0.28    $\pm$       0.19   &       0.42    $\pm$  
     0.29   &         1217 &         1405 &       $2347^{+149 }_{-280 }$ \\
         428 & $        2331^{+         193}_{-         627}$ &     0.09  $\pm$      0.08   &       0.13    $\pm$  
     0.13   &         1774 &         2048 &       $2327^{+195 }_{-669 }$ \\
     1658 & $        2776^{+         26}_{-         28}$ &     0.22  $\pm$      0.02   &       0.33    $\pm$  
     0.02   &         1875 &         2165 &       -- 
     
\enddata
\vspace{0.02cm}
\tablenotetext{a}{Albedos corrected for thermal emission can be found in Table \ref{tab:corralb}.}

\tablenotetext{b}{When the derived error bounds for some of the candidates included zero we did not give a lower limit for the night side temperature.}
\end{deluxetable}

 \begin{deluxetable}{lccc}
\tablecolumns{8}
\tablewidth{0pt}
\tablecaption{Albedos corrected for thermal emission for no $f_{dist}=\frac{1}{2}$ and full redistribution $f_{dist}=\frac{2}{3}$  \label{tab:corralb}}
\tablehead{ \colhead{KOI} & \colhead{$A_g$}&  \colhead{$A_{g,c}(f_{dist}=\frac{1}{2})$}   & \colhead{$A_{g,c}(f_{dist}=\frac{2}{3})$} }

\startdata

     1&    0.05&    0.03&   -0.05\tablenotemark{a} \\
      2&     0.27&     0.23&     0.1\\
      10&     0.11&    0.07&   -0.06\tablenotemark{a}\\
      13&     0.27&    0.09&    -0.27\tablenotemark{a}\\
      17&    0.07&    0.06&   -0.03\tablenotemark{a}\\
      18&     0.16&     0.15&    0.08\\
      20&    0.09&    0.07&   -0.03\tablenotemark{a}\\
      97&     0.32&     0.31&     0.28\\
      127&     0.15&     0.15&     0.13\\
      135&    0.06&   -0.01\tablenotemark{a}&    -0.22\tablenotemark{a}\\
      196&     0.18&     0.16&    0.09\\
      202&     0.11&    0.08&   -0.05\tablenotemark{a}\\
      203&    0.08&    0.05&   -0.08\tablenotemark{a}\\
      204&     0.28&     0.28&     0.26\\
      428&    0.09&    0.02&    -0.21\tablenotemark{a}\\
      1658&     0.22&     0.18&    0.06

\enddata
\vspace{0.02cm}

\tablenotetext{a}{equivalent to a zero albedo}
\end{deluxetable}

\subsection{Non detections}\label{sec:non_detect}

For the following planets from our sample we were not able to detect secondary eclipses.
The upper limits for secondary eclipse depths, albedos, and brightness temperatures of these candidates were calculated by adding the 1 sigma errors from the co-variance matrix to the best fit values.

 \subsubsection{KOI 3.01 / HAT-P-11b / Kepler 3b}

HAT-P-11b (KOI 3.01 or Kepler 3b, \citealt{2010ApJ...710.1724B}) is a Hot Neptune type planet ($17 M_e, 3.8 R_e$) orbiting a bright (V = 9.59) and metal rich K4 dwarf star with a period of 4.89 days. This planet, that was already detected before the start of the \textsl{Kepler} mission is he brightest in our sample and the whole \textsl{Kepler} catalog with a magnitude of 9.174 in the \textsl{Kepler} band. Several other groups already analyzed the phasecurves with no detection of a secondary eclipse (e.g \citealt{2011MNRAS.417.2166S}, \citealt{2011ApJ...740...33D} and \citealt{2011ApJ...743...61S}). Figure \ref{fig:non_detect} (top, left) shows our fit results for KOI 3.01.

The noise in KOI-3.01 is  much bigger than in all other planets of our sample, which misleads our models to an unrealistic albedo  value. However, given the orbital parameters, the maximum secondary eclipse depth we should ever find (assuming it is not self luminous) is 12 ppm. Therefore we did not carry through the temperature calculations for KOI-3.01.

   \subsubsection{KOI 7.01 / Kepler 4b}

Kepler 4b (or KOI 7.01, \citealt{2010ApJ...713L.126B} ) is a $24.5 \pm 3.8 M_e$ and $3.99 \pm 0.21 R_e$ planet with period of 3.21 days around a 4.5 Gyr old near-turnoff G0 star. With a density of about 1.9 $g/cm^3$  Kepler 4b is slightly denser and more massive than Neptune, but about the same size. \cite{2011ApJ...730...50K} exclude a secondary eclipse with an upper limit of 104 ppm for its depth, which corresponds to an upper limit for the brightness temperature of 3988K.

\begin{figure*}
  \centering
      \includegraphics*[width=\textwidth]{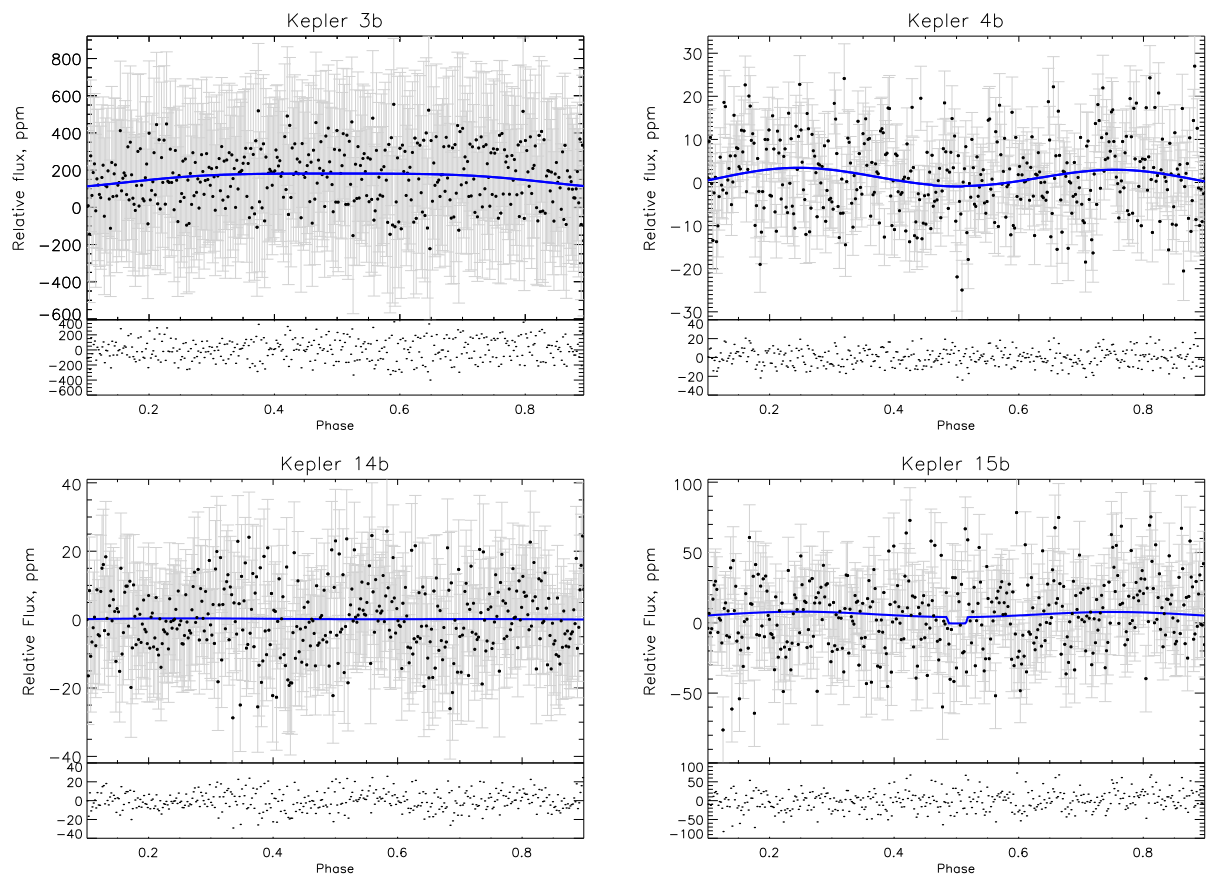}
       \caption{Fitted lightcurves for KOI-3.01/HAT-P-11b/Kepler 3b (top, left)  KOI-7.01/Kepler 4b (top, right), KOI-98.01/Kepler 14b (bottom, left) and KOI-128.01/Kepler 15b (bottom, right).  Blue curves are the best-fit model. None of these systems showed a significant secondary eclipse larger than our noise threshold. However, we were able to calculate upper limits for some of the parameters of these systems (see Table \ref{tab:uplim}).}
     \label{fig:non_detect}
\end{figure*}

Our analysis confirms this and is also consistent with a non-detection to a level of $< 9$ ppm, which enables us to constrain the brightness temperature $T_b$ to $ < 2797$  K. Figure \ref{fig:non_detect} (top, right) and Table \ref{tab:uplim} show our fit results for KOI 7.01.

    \subsubsection{KOI 98.01 / Kepler 14b}
   Kepler 14b (or KOI 98.01) is a 8.40 $M_{Jup}$ and 1.136 $R_{Jup}$ planet on a 6.79 day orbit  around an F star in a binary system \citep{2011ApJS..197....3B}.

   Our analysis, which used an additional polynomial fit to correct for  systematics caused by a close visual binary, is the first of this kind for KOI-98.01. The results show an eclipse depth consistent with a non detection of $<10$ ppm. This leads to a brightness temperature limit $T_b < 2415$  K, a geometric albedo $A_g < 0.17 $. Figure \ref{fig:non_detect} (bottom, left) and Table \ref{tab:uplim} show our fit results for KOI 98.01.

     \subsubsection{KOI 128.01 / Kepler 15b}

  Kepler 15b (\citealt{2011ApJS..197...13E}) is a  $ 0.66 \pm 0.1 M_{Jup}$, $0.96\pm0.06 R_{Jup}$ planet in 4.94 day orbit around a metal-rich ([Fe/H]=0.36 $\pm$ 0.07) G star; its mean density of $0.9 \pm 0.2 g/cm^3$ suggests a significant enrichment in heavy elements. \cite{2011ApJS..197...13E} find no sign of a secondary eclipse.

  For KOI-128.01, we find an eclipse depth consistent with a non-detection of $<11$ ppm. This corresponds to a brightness temperature $T_b$ limit of $< 2039$  K and a geometric albedo limit $A_g$ of $<$ 0.11. Figure \ref{fig:non_detect} (bottom, right) and Table \ref{tab:uplim} show our fit results for KOI 128.01. For KOI-128.01, we used the sum of the (0.5 sigma) detected secondary eclipse value and the error in that value as maximum detectable eclipse value.

 \begin{deluxetable}{lccccc}
\tablecolumns{8}
\tablewidth{0pt}
\tablecaption{Upper limits for planets without detected secondary eclipse. 
 \label{tab:uplim}}
\tablehead{ \colhead{KOI} & \colhead{ecl. depth [ppm]}&  \colhead{$T_b$ [K]}   & \colhead{$A_g$} &\colhead{$A_b$} }\startdata
3\tablenotemark{1} &  $<$ 147  & -- & -- & -- \\
7 & $<$ 9& $<$ 2797 & $<$ 0.62 &$<$ 0.93\\
98 & $<$ 10& $<$  2415& $<$ 0.17 &$<$ 0.26\\
128\tablenotemark{2} & $<$ 11 & $<$ 2039 & $<$ 0.11 &$<$ 0.17
\enddata
\vspace{0.02cm}
\tablenotetext{1}{The noise in KOI-3.01 is  much bigger than in all other planets of our sample, which misleads our models to an unrealistic albedo  value. However, given the orbital parameters, the maximum secondary eclipse depth we should ever find (assuming it is not self luminous) is 12 ppm. Therefore we did not carry through the temperature calculations for KOI-3.01}
\tablenotetext{2}{For KOI-128.01, we used the sum of the (0.5 sigma) detected secondary eclipse value and the error in that value as maximum detectable eclipse value.}
\end{deluxetable}

%
%
%
%

\subsection{Excluded planets}

We also excluded planets from our sample that are part of multiple systems  \citep[e.g.][]{2014ApJ...784...45R}, circumbinary planets \citep[e.g.][]{2011Sci...333.1602D} and highly variable host star systems. Excluded multiple systems containing planets that fall into our sample are KOI 46 / Kepler 10, KOI 137 / Kepler 18, KOI 338 / Kepler 141,  KOI 1779 / Kepler 318 and KOI 1805 / Kepler 319. We excluded KOI 63.01 / Kepler 63b \citep{2013ApJ...775...54S} from our sample because we were not able to apply our methods. In this case the lightcurve was dominated by contributions from an  additional signal with a different periodicity most probably induced by stellar activity.

 \subsection{The case of KOI 200 / Kepler 74b and KOI-2133 / Kepler 91b}

    The planet Kepler 74b \citep{2013AandA...554A.114H} has mass and radius of 0.68   $\pm$   0.09 $M_{Jup}$ and 1.32   $\pm$   0.14 $R_{Jup}$ and orbits its  F8V host star in 7.34 days.
     Kepler 91b \citep{2014AandA...562A.109L} ($ M_p=0.88^{+0.17}_{-0.33} ~M_{Jup}$, $R_p=1.384^{+0.011}_{-0.054} ~R_{Jup}$) orbits its host star ($R_{*}=6.30\pm 0.16 ~R_{Sun}$ , $M_{*}=1.31\pm 0.10 ~ M_{Sun} $) only $1.32^{+0.07}_{-0.22} ~ R_{*}$ away from the stellar atmosphere at the pericenter. \cite{2014AandA...562A.109L} argue that Kepler 91b could therefore be at a stage of the planet engulfment and estimate that Kepler 91b will be swallowed by its host star in less than 55 Myr.
 They derive phasecurve parameters $A_e=121 \pm 33$ ppm, $A_p=25 \pm 15 ppm$ and $A_d = 3 \pm 1$ ppm and no clear secondary eclipse, but 3 other dips in the lightcurve.

\begin{figure}
  \centering
      \includegraphics*[width=\textwidth]{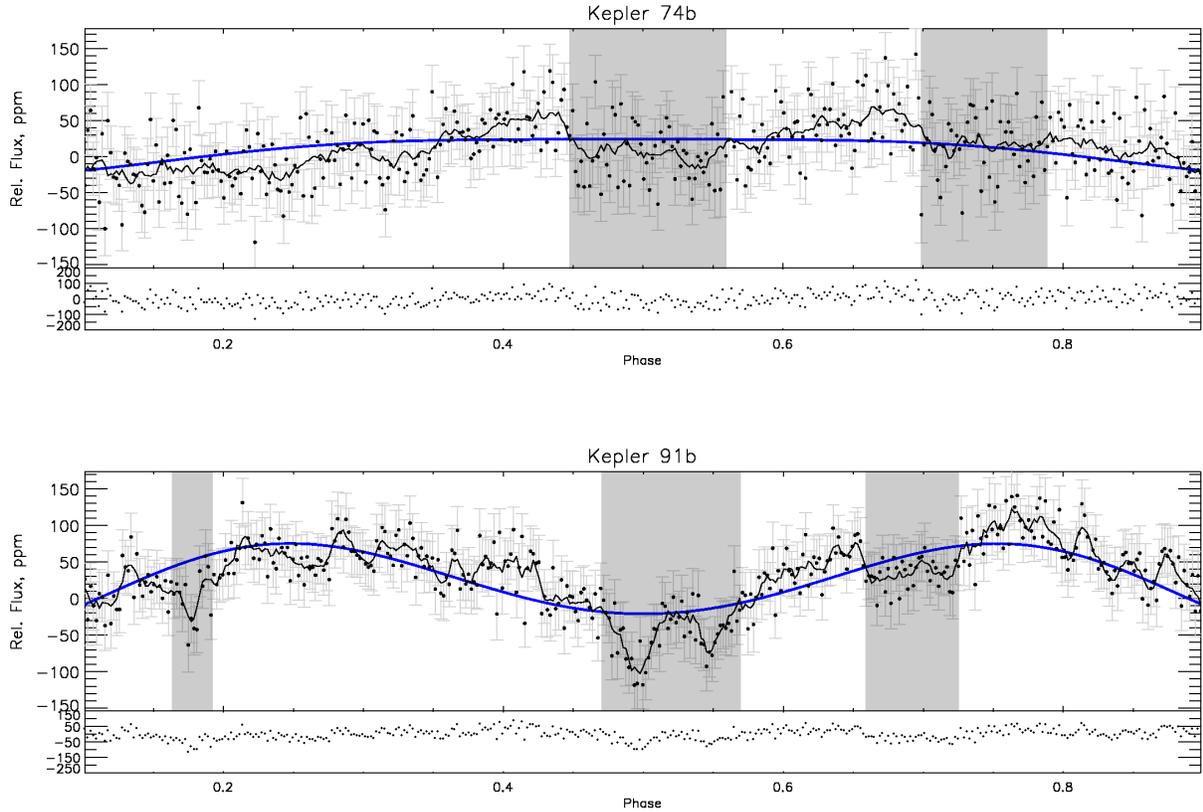}
       \caption{Lightcurves of the systems KOI-200/Kepler 74b (top) and KOI-2133/Kepler 91b (bottom). The grey regions of the lightcurves show extra dimmings that are not explainable with just a secondary eclipse. Our analysis method (blue fits) was not able to recover the secondary eclipse signals from these 'noisy' lightcurves accurately.}
     \label{fig:fig200}
\end{figure}

Our analysis method (see blue fits in Figure \ref{fig:fig200}) was not able to recover the signals from these lightcurves accurately. The reason is that both planets do not show a typical secondary eclipse signature,  but instead also a series of extra dimmings at times other than of the expected eclipse (see grey Figure \ref{fig:fig200}). The timing of some of these extra dips in the lightcurve at 0.166 of the period after transit and/or eclipse may be evidence for the presence of Trojan satellites - however, as also stated in \cite{2014AandA...562A.109L}, this claim needs detailed stability studies to be confirmed. Another explanation is correlated noise and/or stellar activity dominating these lightcurves. We were not able to recover any useful results for the secondary eclipse depths and most of the other phase curve parameters and therefore were not able to compare to previous results and excluded these two targets from our analysis.

\section{Summary and discussion}

With our consistent analysis we were able to confirm and in most cases improve parameters derived by previous studies. We present new results for Kepler 1b-8b, 12b-15b, 17b, 40b, 41b, 43b, 44b, 76b, 77b, and 412b.

\subsection{Comparison to other fitting routines - future improvements}

For the cases of previously analyzed targets we were able to confirm (within less than 2 $\sigma$ for almost all cases) results derived from various previous or parallel analyses that used very different modeling approaches, from relatively simple boxcar fits (without a phase curve model) of only the secondary eclipse to very sophisticated Bayesian codes fitting all system parameters in an integrated way. The fact that we reproduce most of these results demonstrates the value of our compromise approach to combine a state-of-the-art phasecurve model including all important contributions with a robust least-squares fit to trade off between number of systems and computing time. Also our goal was to focus on eclipses and phasecurves while fixing all other parameters to previously derived values. This significantly reduces the number of potential degeneracies that usually call for these more elaborated methods.

 However, in order to increase the statistical relevance of our results we are currently working to extend the analysis to the whole sample of 489 Kepler Objects of Interest with $R_p > 4 R_e$ , $P < 10d$, $V_{mag} < 15$: we plan to apply EXONEST \citep{2014ApJ...795..112P}, a Bayesian model selection algorithm to the whole set of 489 candidates. With a sample of that size we hope to find statistically significant correlations of stellar and planetary parameters with the position of the planet, e.g., in albedo $vs$ incoming flux phase space (see Figure \ref{fig:corr_star}).

\subsection{Correlations with system parameters}
\begin{figure}[t!]
  \centering
      \includegraphics*[width=\textwidth]{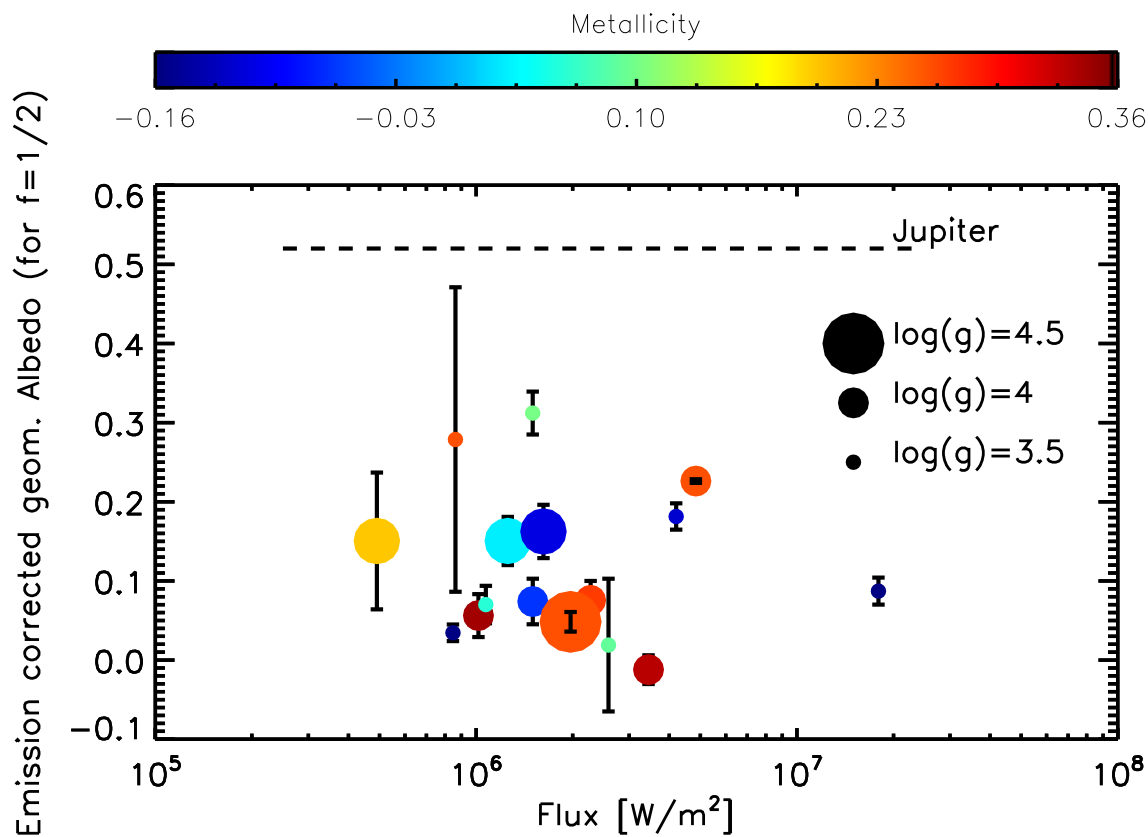}
       \caption{Emission-corrected geometric albedo $A_{g,c}$ for $(f=1/2)$ versus the incident stellar flux for our sample of Kepler giant planets (compare to Fig. 1 in \citealt{2013ApJ...777..100H}). The colors represent the host star's metallicity [Fe/H] (see colorbar) and the size of the symbol corresponds to the host's  surface gravity $log(g)$ (see legend). There is no obvious correlation with the distribution. The dashed line represents Jupiters's albedo of 0.52 for comparison; all of the Hot Jupiters in our study have lower albedos.}
     \label{fig:corr_star}
\end{figure}

For the following analysis of a potential correlation of the albedo with stellar or planetary parameters (see e.g. Figure \ref{fig:corr_star}) we used the albedo values, that were corrected for thermal emission (assuming no redistribution; $f_{dist}=1/2$) that are shown in Table \ref{tab:corralb} (center).

Our results confirm the general trend of relatively low albedos for most of the Hot Jupiters, but we also show outliers with higher albedos.

We see no significant correlations in our data. Neither the stellar parameters ($[Fe/H]$ and $log(g)$, see Figure \ref{fig:corr_star}) nor the planetary characteristics (mass, radius, density and surface gravity) correlate with the derived parameters. When excluding the planets with large errors in the albedo, there are indications that massive planets, very dense and very bloated planets tend to be low in albedo -- i.e., density extremes produce low albedos.

Even though we present a relatively large sample characterized in this comprehensive and consistent manner, our sample size is still too small to draw significant statistical conclusions. Future efforts will include analyzing all of the candidates as well as the complete Kepler dataset of 18 available quarters (Q0-Q17).

With \textsl{TESS} \citep{2010AAS...21545006R} and \textsl{PLATO} \citep{2013arXiv1310.0696R,2015arXiv150303251H} on the horizon the future will bear an even bigger data set, marking great potential for characterizations in a similar way. Such future observations and analyses will include many more planets in a similar range of planetary and stellar parameters. A comparative analysis beyond their basic parameters of a large number of planets orbiting a variety of host stars  will probe and eventually solve the fundamental underlying questions on planet formation, migration and composition.


%

\acknowledgments

We want to thank Martin Still (Director of the Kepler Guest Observer Office) for his valuable comments and frequent support on the use of PyKE. We thank Avi Shporer, Kevin Heng and the anonymous referee for valuable comments on the manuscript.

This research was supported by internal funds at the Rensselaer Polytechnic Institute.
DA's research was also supported by
an appointment to the NASA Postdoctoral Program at the Goddard Space Flight Center, administered by Oak Ridge Associated Universities through a contract with NASA.



{\it Facilities:} \facility{Kepler}.



\appendix





\clearpage


\begin{thebibliography}{}
\expandafter\ifx\csname natexlab\endcsname\relax\def\natexlab#1{#1}\fi

\bibitem[{{Bakos} {et~al.}(2010){Bakos}, {Torres}, {P{\'a}l}, {Hartman},
  {Kov{\'a}cs}, {Noyes}, {Latham}, {Sasselov}, {Sip{\H o}cz}, {Esquerdo},
  {Fischer}, {Johnson}, {Marcy}, {Butler}, {Isaacson}, {Howard}, {Vogt},
  {Kov{\'a}cs}, {Fernandez}, {Mo{\'o}r}, {Stefanik}, {L{\'a}z{\'a}r}, {Papp},
  \& {S{\'a}ri}}]{2010ApJ...710.1724B}
{Bakos}, G.~{\'A}., {Torres}, G., {P{\'a}l}, A., {et~al.} 2010, \apj, 710, 1724

\bibitem[{{Ballard} {et~al.}(2011){Ballard}, {Fabrycky}, {Fressin},
  {Charbonneau}, {Desert}, {Torres}, {Marcy}, {Burke}, {Isaacson}, {Henze},
  {Steffen}, {Ciardi}, {Howell}, {Cochran}, {Endl}, {Bryson}, {Rowe}, {Holman},
  {Lissauer}, {Jenkins}, {Still}, {Ford}, {Christiansen}, {Middour}, {Haas},
  {Li}, {Hall}, {McCauliff}, {Batalha}, {Koch}, \&
  {Borucki}}]{2011ApJ...743..200B}
{Ballard}, S., {Fabrycky}, D., {Fressin}, F., {et~al.} 2011, \apj, 743, 200

\bibitem[{{Barclay} {et~al.}(2012){Barclay}, {Huber}, {Rowe}, {Fortney},
  {Morley}, {Quintana}, {Fabrycky}, {Barentsen}, {Bloemen}, {Christiansen},
  {Demory}, {Fulton}, {Jenkins}, {Mullally}, {Ragozzine}, {Seader}, {Shporer},
  {Tenenbaum}, \& {Thompson}}]{2012ApJ...761...53B}
{Barclay}, T., {Huber}, D., {Rowe}, J.~F., {et~al.} 2012, \apj, 761, 53

\bibitem[{{Barclay} {et~al.}(2013){Barclay}, {Burke}, {Howell}, {Rowe},
  {Huber}, {Isaacson}, {Jenkins}, {Kolbl}, {Marcy}, {Quintana}, {Still},
  {Twicken}, {Bryson}, {Borucki}, {Caldwell}, {Ciardi}, {Clarke},
  {Christiansen}, {Coughlin}, {Fischer}, {Li}, {Haas}, {Hunter}, {Lissauer},
  {Mullally}, {Sabale}, {Seader}, {Smith}, {Tenenbaum}, {Kamal Uddin}, \&
  {Thompson}}]{2013ApJ...768..101B}
{Barclay}, T., {Burke}, C.~J., {Howell}, S.~B., {et~al.} 2013, \apj, 768, 101

\bibitem[{{Batalha} {et~al.}(2011){Batalha}, {Borucki}, {Bryson}, {Buchhave},
  {Caldwell}, {Christensen-Dalsgaard}, {Ciardi}, {Dunham}, {Fressin},
  {Gautier}, {Gilliland}, {Haas}, {Howell}, {Jenkins}, {Kjeldsen}, {Koch},
  {Latham}, {Lissauer}, {Marcy}, {Rowe}, {Sasselov}, {Seager}, {Steffen},
  {Torres}, {Basri}, {Brown}, {Charbonneau}, {Christiansen}, {Clarke},
  {Cochran}, {Dupree}, {Fabrycky}, {Fischer}, {Ford}, {Fortney}, {Girouard},
  {Holman}, {Johnson}, {Isaacson}, {Klaus}, {Machalek}, {Moorehead},
  {Morehead}, {Ragozzine}, {Tenenbaum}, {Twicken}, {Quinn}, {VanCleve},
  {Walkowicz}, {Welsh}, {Devore}, \& {Gould}}]{2011ApJ...729...27B}
{Batalha}, N.~M., {Borucki}, W.~J., {Bryson}, S.~T., {et~al.} 2011, \apj, 729,
  27

\bibitem[{{Batalha} {et~al.}(2013){Batalha}, {Rowe}, {Bryson}, {Barclay},
  {Burke}, {Caldwell}, {Christiansen}, {Mullally}, {Thompson}, {Brown},
  {Dupree}, {Fabrycky}, {Ford}, {Fortney}, {Gilliland}, {Isaacson}, {Latham},
  {Marcy}, {Quinn}, {Ragozzine}, {Shporer}, {Borucki}, {Ciardi}, {Gautier},
  {Haas}, {Jenkins}, {Koch}, {Lissauer}, {Rapin}, {Basri}, {Boss}, {Buchhave},
  {Carter}, {Charbonneau}, {Christensen-Dalsgaard}, {Clarke}, {Cochran},
  {Demory}, {Desert}, {Devore}, {Doyle}, {Esquerdo}, {Everett}, {Fressin},
  {Geary}, {Girouard}, {Gould}, {Hall}, {Holman}, {Howard}, {Howell},
  {Ibrahim}, {Kinemuchi}, {Kjeldsen}, {Klaus}, {Li}, {Lucas}, {Meibom},
  {Morris}, {Pr{\v s}a}, {Quintana}, {Sanderfer}, {Sasselov}, {Seader},
  {Smith}, {Steffen}, {Still}, {Stumpe}, {Tarter}, {Tenenbaum}, {Torres},
  {Twicken}, {Uddin}, {Van Cleve}, {Walkowicz}, \&
  {Welsh}}]{2013ApJS..204...24B}
{Batalha}, N.~M., {Rowe}, J.~F., {Bryson}, S.~T., {et~al.} 2013, \apjs, 204, 24

\bibitem[{{Bonomo} {et~al.}(2012){Bonomo}, {H{\'e}brard}, {Santerne}, {Santos},
  {Deleuil}, {Almenara}, {Bouchy}, {D{\'{\i}}az}, {Moutou}, \&
  {Vanhuysse}}]{2012AandA...538A..96B}
{Bonomo}, A.~S., {H{\'e}brard}, G., {Santerne}, A., {et~al.} 2012, \aap, 538,
  A96

\bibitem[{{Borucki} {et~al.}(2009){Borucki}, {Koch}, {Jenkins}, {Sasselov},
  {Gilliland}, {Batalha}, {Latham}, {Caldwell}, {Basri}, {Brown},
  {Christensen-Dalsgaard}, {Cochran}, {DeVore}, {Dunham}, {Dupree}, {Gautier},
  {Geary}, {Gould}, {Howell}, {Kjeldsen}, {Lissauer}, {Marcy}, {Meibom},
  {Morrison}, \& {Tarter}}]{2009Sci...325..709B}
{Borucki}, W.~J., {Koch}, D., {Jenkins}, J., {et~al.} 2009, Science, 325, 709

\bibitem[{{Borucki} {et~al.}(2010{\natexlab{a}}){Borucki}, {Koch}, {Brown},
  {Basri}, {Batalha}, {Caldwell}, {Cochran}, {Dunham}, {Gautier}, {Geary},
  {Gilliland}, {Howell}, {Jenkins}, {Latham}, {Lissauer}, {Marcy}, {Monet},
  {Rowe}, \& {Sasselov}}]{2010ApJ...713L.126B}
{Borucki}, W.~J., {Koch}, D.~G., {Brown}, T.~M., {et~al.} 2010{\natexlab{a}},
  \apjl, 713, L126

\bibitem[{{Borucki} {et~al.}(2010{\natexlab{b}}){Borucki}, {Koch}, {Basri},
  {Batalha}, {Brown}, {Caldwell}, {Caldwell}, {Christensen-Dalsgaard},
  {Cochran}, {DeVore}, {Dunham}, {Dupree}, {Gautier}, {Geary}, {Gilliland},
  {Gould}, {Howell}, {Jenkins}, {Kondo}, {Latham}, {Marcy}, {Meibom},
  {Kjeldsen}, {Lissauer}, {Monet}, {Morrison}, {Sasselov}, {Tarter}, {Boss},
  {Brownlee}, {Owen}, {Buzasi}, {Charbonneau}, {Doyle}, {Fortney}, {Ford},
  {Holman}, {Seager}, {Steffen}, {Welsh}, {Rowe}, {Anderson}, {Buchhave},
  {Ciardi}, {Walkowicz}, {Sherry}, {Horch}, {Isaacson}, {Everett}, {Fischer},
  {Torres}, {Johnson}, {Endl}, {MacQueen}, {Bryson}, {Dotson}, {Haas},
  {Kolodziejczak}, {Van Cleve}, {Chandrasekaran}, {Twicken}, {Quintana},
  {Clarke}, {Allen}, {Li}, {Wu}, {Tenenbaum}, {Verner}, {Bruhweiler}, {Barnes},
  \& {Prsa}}]{2010Sci...327..977B}
{Borucki}, W.~J., {Koch}, D., {Basri}, G., {et~al.} 2010{\natexlab{b}},
  Science, 327, 977

\bibitem[{{Borucki} {et~al.}(2011){Borucki}, {Koch}, {Basri}, {Batalha},
  {Brown}, {Bryson}, {Caldwell}, {Christensen-Dalsgaard}, {Cochran}, {DeVore},
  {Dunham}, {Gautier}, {Geary}, {Gilliland}, {Gould}, {Howell}, {Jenkins},
  {Latham}, {Lissauer}, {Marcy}, {Rowe}, {Sasselov}, {Boss}, {Charbonneau},
  {Ciardi}, {Doyle}, {Dupree}, {Ford}, {Fortney}, {Holman}, {Seager},
  {Steffen}, {Tarter}, {Welsh}, {Allen}, {Buchhave}, {Christiansen}, {Clarke},
  {Das}, {D{\'e}sert}, {Endl}, {Fabrycky}, {Fressin}, {Haas}, {Horch},
  {Howard}, {Isaacson}, {Kjeldsen}, {Kolodziejczak}, {Kulesa}, {Li}, {Lucas},
  {Machalek}, {McCarthy}, {MacQueen}, {Meibom}, {Miquel}, {Prsa}, {Quinn},
  {Quintana}, {Ragozzine}, {Sherry}, {Shporer}, {Tenenbaum}, {Torres},
  {Twicken}, {Van Cleve}, {Walkowicz}, {Witteborn}, \&
  {Still}}]{2011ApJ...736...19B}
{Borucki}, W.~J., {Koch}, D.~G., {Basri}, G., {et~al.} 2011, \apj, 736, 19

\bibitem[{{Borucki} {et~al.}(2013){Borucki}, {Agol}, {Fressin}, {Kaltenegger},
  {Rowe}, {Isaacson}, {Fischer}, {Batalha}, {Lissauer}, {Marcy}, {Fabrycky},
  {D{\'e}sert}, {Bryson}, {Barclay}, {Bastien}, {Boss}, {Brugamyer},
  {Buchhave}, {Burke}, {Caldwell}, {Carter}, {Charbonneau}, {Crepp},
  {Christensen-Dalsgaard}, {Christiansen}, {Ciardi}, {Cochran}, {DeVore},
  {Doyle}, {Dupree}, {Endl}, {Everett}, {Ford}, {Fortney}, {Gautier}, {Geary},
  {Gould}, {Haas}, {Henze}, {Howard}, {Howell}, {Huber}, {Jenkins}, {Kjeldsen},
  {Kolbl}, {Kolodziejczak}, {Latham}, {Lee}, {Lopez}, {Mullally}, {Orosz},
  {Prsa}, {Quintana}, {Sanchis-Ojeda}, {Sasselov}, {Seader}, {Shporer},
  {Steffen}, {Still}, {Tenenbaum}, {Thompson}, {Torres}, {Twicken}, {Welsh}, \&
  {Winn}}]{2013Sci...340..587B}
{Borucki}, W.~J., {Agol}, E., {Fressin}, F., {et~al.} 2013, Science, 340, 587

\bibitem[{{Buchhave} {et~al.}(2011){Buchhave}, {Latham}, {Carter},
  {D{\'e}sert}, {Torres}, {Adams}, {Bryson}, {Charbonneau}, {Ciardi}, {Kulesa},
  {Dupree}, {Fischer}, {Fressin}, {Gautier}, {Gilliland}, {Howell}, {Isaacson},
  {Jenkins}, {Marcy}, {McCarthy}, {Rowe}, {Batalha}, {Borucki}, {Brown},
  {Caldwell}, {Christiansen}, {Cochran}, {Deming}, {Dunham}, {Everett}, {Ford},
  {Fortney}, {Geary}, {Girouard}, {Haas}, {Holman}, {Horch}, {Klaus},
  {Knutson}, {Koch}, {Kolodziejczak}, {Lissauer}, {Machalek}, {Mullally},
  {Still}, {Quinn}, {Seager}, {Thompson}, \& {Van Cleve}}]{2011ApJS..197....3B}
{Buchhave}, L.~A., {Latham}, D.~W., {Carter}, J.~A., {et~al.} 2011, \apjs, 197,
  3

\bibitem[{{Burke} {et~al.}(2014){Burke}, {Bryson}, {Mullally}, {Rowe},
  {Christiansen}, {Thompson}, {Coughlin}, {Haas}, {Batalha}, {Caldwell},
  {Jenkins}, {Still}, {Barclay}, {Borucki}, {Chaplin}, {Ciardi}, {Clarke},
  {Cochran}, {Demory}, {Esquerdo}, {Gautier}, {Gilliland}, {Girouard}, {Havel},
  {Henze}, {Howell}, {Huber}, {Latham}, {Li}, {Morehead}, {Morton}, {Pepper},
  {Quintana}, {Ragozzine}, {Seader}, {Shah}, {Shporer}, {Tenenbaum}, {Twicken},
  \& {Wolfgang}}]{2014ApJS..210...19B}
{Burke}, C.~J., {Bryson}, S.~T., {Mullally}, F., {et~al.} 2014, \apjs, 210, 19

\bibitem[Burke et al.(2015)]{2015arXiv150604175B} Burke, C.~J., 
Christiansen, J.~L., Mullally, F., et al.\ 2015, arXiv:1506.04175 

\bibitem[{{Claret} \& {Bloemen}(2011)}]{2011AandA...529A..75C}
{Claret}, A., \& {Bloemen}, S. 2011, \aap, 529, A75

\bibitem[{{Coughlin} \& {L{\'o}pez-Morales}(2012)}]{2012AJ....143...39C}
{Coughlin}, J.~L., \& {L{\'o}pez-Morales}, M. 2012, \aj, 143, 39

\bibitem[{{Cowan} \& {Agol}(2011)}]{2011ApJ...729...54C}
{Cowan}, N.~B., \& {Agol}, E. 2011, \apj, 729, 54

\bibitem[{{Deleuil} {et~al.}(2014){Deleuil}, {Almenara}, {Santerne}, {Barros},
  {Havel}, {H{\'e}brard}, {Bonomo}, {Bouchy}, {Bruno}, {Damiani},
  {D{\'{\i}}az}, {Montagnier}, \& {Moutou}}]{2014arXiv1401.6811D}
{Deleuil}, M., {Almenara}, J.-M., {Santerne}, A., {et~al.} 2014, ArXiv
  e-prints, arXiv:1401.6811

\bibitem[{{Deming} {et~al.}(2011){Deming}, {Sada}, {Jackson}, {Peterson},
  {Agol}, {Knutson}, {Jennings}, {Haase}, \& {Bays}}]{2011ApJ...740...33D}
{Deming}, D., {Sada}, P.~V., {Jackson}, B., {et~al.} 2011, \apj, 740, 33

\bibitem[{{Demory} \& {Seager}(2011)}]{2011IAUS..276..475D}
{Demory}, B.-O., \& {Seager}, S. 2011, in IAU Symposium, Vol. 276, IAU
  Symposium, ed. A.~{Sozzetti}, M.~G. {Lattanzi}, \& A.~P. {Boss}, 475--476

\bibitem[{{Demory} {et~al.}(2011){Demory}, {Seager}, {Madhusudhan}, {Kjeldsen},
  {Christensen-Dalsgaard}, {Gillon}, {Rowe}, {Welsh}, {Adams}, {Dupree},
  {McCarthy}, {Kulesa}, {Borucki}, \& {Koch}}]{2011ApJ...735L..12D}
{Demory}, B.-O., {Seager}, S., {Madhusudhan}, N., {et~al.} 2011, \apjl, 735,
  L12

\bibitem[{{Demory} {et~al.}(2013){Demory}, {de Wit}, {Lewis}, {Fortney},
  {Zsom}, {Seager}, {Knutson}, {Heng}, {Madhusudhan}, {Gillon}, {Barclay},
  {Desert}, {Parmentier}, \& {Cowan}}]{2013ApJ...776L..25D}
{Demory}, B.-O., {de Wit}, J., {Lewis}, N., {et~al.} 2013, \apjl, 776, L25

\bibitem[{{D{\'e}sert} {et~al.}(2011){D{\'e}sert}, {Charbonneau}, {Fortney},
  {Madhusudhan}, {Knutson}, {Fressin}, {Deming}, {Borucki}, {Brown},
  {Caldwell}, {Ford}, {Gilliland}, {Latham}, {Marcy}, \&
  {Seager}}]{2011ApJS..197...11D}
{D{\'e}sert}, J.-M., {Charbonneau}, D., {Fortney}, J.~J., {et~al.} 2011, \apjs,
  197, 11

\bibitem[{{Doyle} {et~al.}(2011){Doyle}, {Carter}, {Fabrycky}, {Slawson},
  {Howell}, {Winn}, {Orosz}, {Prsa}, {Welsh}, {Quinn}, {Latham}, {Torres},
  {Buchhave}, {Marcy}, {Fortney}, {Shporer}, {Ford}, {Lissauer}, {Ragozzine},
  {Rucker}, {Batalha}, {Jenkins}, {Borucki}, {Koch}, {Middour}, {Hall},
  {McCauliff}, {Fanelli}, {Quintana}, {Holman}, {Caldwell}, {Still},
  {Stefanik}, {Brown}, {Esquerdo}, {Tang}, {Furesz}, {Geary}, {Berlind},
  {Calkins}, {Short}, {Steffen}, {Sasselov}, {Dunham}, {Cochran}, {Boss},
  {Haas}, {Buzasi}, \& {Fischer}}]{2011Sci...333.1602D}
{Doyle}, L.~R., {Carter}, J.~A., {Fabrycky}, D.~C., {et~al.} 2011, Science,
  333, 1602

\bibitem[{{Dressing} \& {Charbonneau}(2013)}]{2013ApJ...767...95D}
{Dressing}, C.~D., \& {Charbonneau}, D. 2013, \apj, 767, 95

\bibitem[{{Dunham} {et~al.}(2010){Dunham}, {Borucki}, {Koch}, {Batalha},
  {Buchhave}, {Brown}, {Caldwell}, {Cochran}, {Endl}, {Fischer}, {F{\H
  u}r{\'e}sz}, {Gautier}, {Geary}, {Gilliland}, {Gould}, {Howell}, {Jenkins},
  {Kjeldsen}, {Latham}, {Lissauer}, {Marcy}, {Meibom}, {Monet}, {Rowe}, \&
  {Sasselov}}]{2010ApJ...713L.136D}
{Dunham}, E.~W., {Borucki}, W.~J., {Koch}, D.~G., {et~al.} 2010, \apjl, 713,
  L136

\bibitem[{{Endl} {et~al.}(2011){Endl}, {MacQueen}, {Cochran}, {Brugamyer},
  {Buchhave}, {Rowe}, {Lucas}, {Isaacson}, {Bryson}, {Howell}, {Fortney},
  {Hansen}, {Borucki}, {Caldwell}, {Christiansen}, {Ciardi}, {Demory},
  {Everett}, {Ford}, {Haas}, {Holman}, {Horch}, {Jenkins}, {Koch}, {Lissauer},
  {Machalek}, {Still}, {Welsh}, {Sanderfer}, {Seader}, {Smith}, {Thompson}, \&
  {Twicken}}]{2011ApJS..197...13E}
{Endl}, M., {MacQueen}, P.~J., {Cochran}, W.~D., {et~al.} 2011, \apjs, 197, 13

\bibitem[{{Esteves} {et~al.}(2013){Esteves}, {De Mooij}, \&
  {Jayawardhana}}]{2013ApJ...772...51E}
{Esteves}, L.~J., {De Mooij}, E.~J.~W., \& {Jayawardhana}, R. 2013, \apj, 772,
  51

\bibitem[{{Faigler} \& {Mazeh}(2011)}]{2011MNRAS.415.3921F}
{Faigler}, S., \& {Mazeh}, T. 2011, \mnras, 415, 3921

\bibitem[{{Faigler} {et~al.}(2013){Faigler}, {Tal-Or}, {Mazeh}, {Latham}, \&
  {Buchhave}}]{2013ApJ...771...26F}
{Faigler}, S., {Tal-Or}, L., {Mazeh}, T., {Latham}, D.~W., \& {Buchhave}, L.~A.
  2013, \apj, 771, 26

\bibitem[{{Fortney} {et~al.}(2011){Fortney}, {Demory}, {D{\'e}sert}, {Rowe},
  {Marcy}, {Isaacson}, {Buchhave}, {Ciardi}, {Gautier}, {Batalha}, {Caldwell},
  {Bryson}, {Nutzman}, {Jenkins}, {Howard}, {Charbonneau}, {Knutson}, {Howell},
  {Everett}, {Fressin}, {Deming}, {Borucki}, {Brown}, {Ford}, {Gilliland},
  {Latham}, {Miller}, {Seager}, {Fischer}, {Koch}, {Lissauer}, {Haas}, {Still},
  {Lucas}, {Gillon}, {Christiansen}, \& {Geary}}]{2011ApJS..197....9F}
{Fortney}, J.~J., {Demory}, B.-O., {D{\'e}sert}, J.-M., {et~al.} 2011, \apjs,
  197, 9

\bibitem[{{Fressin} {et~al.}(2013){Fressin}, {Torres}, {Charbonneau}, {Bryson},
  {Christiansen}, {Dressing}, {Jenkins}, {Walkowicz}, \&
  {Batalha}}]{2013ApJ...766...81F}
{Fressin}, F., {Torres}, G., {Charbonneau}, D., {et~al.} 2013, \apj, 766, 81

\bibitem[{{Gandolfi} {et~al.}(2013){Gandolfi}, {Parviainen}, {Fridlund},
  {Hatzes}, {Deeg}, {Frasca}, {Lanza}, {Prada Moroni}, {Tognelli}, {McQuillan},
  {Aigrain}, {Alonso}, {Antoci}, {Cabrera}, {Carone}, {Csizmadia}, {Djupvik},
  {Guenther}, {Jessen-Hansen}, {Ofir}, \& {Telting}}]{2013AandA...557A..74G}
{Gandolfi}, D., {Parviainen}, H., {Fridlund}, M., {et~al.} 2013, \aap, 557, A74

\bibitem[{{Groot}(2012)}]{2012ApJ...745...55G}
{Groot}, P.~J. 2012, \apj, 745, 55

\bibitem[{{H{\'e}brard} {et~al.}(2013){H{\'e}brard}, {Almenara}, {Santerne},
  {Deleuil}, {Damiani}, {Bonomo}, {Bouchy}, {Bruno}, {D{\'{\i}}az},
  {Montagnier}, \& {Moutou}}]{2013AandA...554A.114H}
{H{\'e}brard}, G., {Almenara}, J.-M., {Santerne}, A., {et~al.} 2013, \aap, 554,
  A114

\bibitem[{{Heng} \& {Demory}(2013)}]{2013ApJ...777..100H}
{Heng}, K., \& {Demory}, B.-O. 2013, \apj, 777, 100

\bibitem[{{Henry} {et~al.}(2000){Henry}, {Marcy}, {Butler}, \&
  {Vogt}}]{2000ApJ...529L..41H}
{Henry}, G.~W., {Marcy}, G.~W., {Butler}, R.~P., \& {Vogt}, S.~S. 2000, \apjl,
  529, L41
  
  \bibitem[Hippke 
\& Angerhausen(2015)]{2015arXiv150800427H} Hippke, M., \& Angerhausen, D.\ 2015, arXiv:1508.00427 

\bibitem[Hippke 
\& Angerhausen(2015)]{2015arXiv150303251H} Hippke, M., \& Angerhausen, D.\ 2015, arXiv:1503.03251 

\bibitem[{{Holman} {et~al.}(2010){Holman}, {Fabrycky}, {Ragozzine}, {Ford},
  {Steffen}, {Welsh}, {Lissauer}, {Latham}, {Marcy}, {Walkowicz}, {Batalha},
  {Jenkins}, {Rowe}, {Cochran}, {Fressin}, {Torres}, {Buchhave}, {Sasselov},
  {Borucki}, {Koch}, {Basri}, {Brown}, {Caldwell}, {Charbonneau}, {Dunham},
  {Gautier}, {Geary}, {Gilliland}, {Haas}, {Howell}, {Ciardi}, {Endl},
  {Fischer}, {F{\"u}r{\'e}sz}, {Hartman}, {Isaacson}, {Johnson}, {MacQueen},
  {Moorhead}, {Morehead}, \& {Orosz}}]{2010Sci...330...51H}
{Holman}, M.~J., {Fabrycky}, D.~C., {Ragozzine}, D., {et~al.} 2010, Science,
  330, 51

\bibitem[{{Howard} {et~al.}(2012){Howard}, {Marcy}, {Bryson}, {Jenkins},
  {Rowe}, {Batalha}, {Borucki}, {Koch}, {Dunham}, {Gautier}, {Van Cleve},
  {Cochran}, {Latham}, {Lissauer}, {Torres}, {Brown}, {Gilliland}, {Buchhave},
  {Caldwell}, {Christensen-Dalsgaard}, {Ciardi}, {Fressin}, {Haas}, {Howell},
  {Kjeldsen}, {Seager}, {Rogers}, {Sasselov}, {Steffen}, {Basri},
  {Charbonneau}, {Christiansen}, {Clarke}, {Dupree}, {Fabrycky}, {Fischer},
  {Ford}, {Fortney}, {Tarter}, {Girouard}, {Holman}, {Johnson}, {Klaus},
  {Machalek}, {Moorhead}, {Morehead}, {Ragozzine}, {Tenenbaum}, {Twicken},
  {Quinn}, {Isaacson}, {Shporer}, {Lucas}, {Walkowicz}, {Welsh}, {Boss},
  {Devore}, {Gould}, {Smith}, {Morris}, {Prsa}, {Morton}, {Still}, {Thompson},
  {Mullally}, {Endl}, \& {MacQueen}}]{2012ApJS..201...15H}
{Howard}, A.~W., {Marcy}, G.~W., {Bryson}, S.~T., {et~al.} 2012, \apjs, 201, 15

\bibitem[{{Jenkins} {et~al.}(2010){Jenkins}, {Borucki}, {Koch}, {Marcy},
  {Cochran}, {Welsh}, {Basri}, {Batalha}, {Buchhave}, {Brown}, {Caldwell},
  {Dunham}, {Endl}, {Fischer}, {Gautier}, {Geary}, {Gilliland}, {Howell},
  {Isaacson}, {Johnson}, {Latham}, {Lissauer}, {Monet}, {Rowe}, {Sasselov},
  {Howard}, {MacQueen}, {Orosz}, {Chandrasekaran}, {Twicken}, {Bryson},
  {Quintana}, {Clarke}, {Li}, {Allen}, {Tenenbaum}, {Wu}, {Meibom}, {Klaus},
  {Middour}, {Cote}, {McCauliff}, {Girouard}, {Gunter}, {Wohler}, {Hall},
  {Ibrahim}, {Kamal Uddin}, {Wu}, {Bhavsar}, {Van Cleve}, {Pletcher}, {Dotson},
  \& {Haas}}]{2010ApJ...724.1108J}
{Jenkins}, J.~M., {Borucki}, W.~J., {Koch}, D.~G., {et~al.} 2010, \apj, 724,
  1108
  
  
\bibitem[Jenkins et al.(2015)]{2015AJ....150...56J} Jenkins, J.~M., 
Twicken, J.~D., Batalha, N.~M., et al.\ 2015, \aj, 150, 56 

\bibitem[{{Kepler Mission Team}(2009)}]{2009yCat.5133....0K}
{Kepler Mission Team}. 2009, VizieR Online Data Catalog, 5133, 0

\bibitem[{{Kipping} \& {Bakos}(2011)}]{2011ApJ...730...50K}
{Kipping}, D., \& {Bakos}, G. 2011, \apj, 730, 50

\bibitem[{{Kipping} {et~al.}(2013){Kipping}, {Hartman}, {Buchhave}, {Schmitt},
  {Bakos}, \& {Nesvorn{\'y}}}]{2013ApJ...770..101K}
{Kipping}, D.~M., {Hartman}, J., {Buchhave}, L.~A., {et~al.} 2013, \apj, 770,
  101

\bibitem[{{Kipping} \& {Spiegel}(2011)}]{2011MNRAS.417L..88K}
{Kipping}, D.~M., \& {Spiegel}, D.~S. 2011, \mnras, 417, L88

\bibitem[Kipping(2010)]{2010MNRAS.408.1758K} Kipping, D.~M.\ 2010, \mnras, 
408, 1758 


\bibitem[{{Koch} {et~al.}(2010){Koch}, {Borucki}, {Rowe}, {Batalha}, {Brown},
  {Caldwell}, {Caldwell}, {Cochran}, {DeVore}, {Dunham}, {Dupree}, {Gautier},
  {Geary}, {Gilliland}, {Howell}, {Jenkins}, {Latham}, {Lissauer}, {Marcy},
  {Morrison}, \& {Tarter}}]{2010ApJ...713L.131K}
{Koch}, D.~G., {Borucki}, W.~J., {Rowe}, J.~F., {et~al.} 2010, \apjl, 713, L131

\bibitem[{{Latham} {et~al.}(2010){Latham}, {Borucki}, {Koch}, {Brown},
  {Buchhave}, {Basri}, {Batalha}, {Caldwell}, {Cochran}, {Dunham}, {F{\H
  u}r{\'e}sz}, {Gautier}, {Geary}, {Gilliland}, {Howell}, {Jenkins},
  {Lissauer}, {Marcy}, {Monet}, {Rowe}, \& {Sasselov}}]{2010ApJ...713L.140L}
{Latham}, D.~W., {Borucki}, W.~J., {Koch}, D.~G., {et~al.} 2010, \apjl, 713,
  L140

\bibitem[{{Lillo-Box} {et~al.}(2014){Lillo-Box}, {Barrado}, {Moya},
  {Montesinos}, {Montalb{\'a}n}, {Bayo}, {Barbieri}, {R{\'e}gulo}, {Mancini},
  {Bouy}, \& {Henning}}]{2014AandA...562A.109L}
{Lillo-Box}, J., {Barrado}, D., {Moya}, A., {et~al.} 2014, \aap, 562, A109

\bibitem[{{Line} {et~al.}(2010){Line}, {Liang}, \&
  {Yung}}]{2010ApJ...717..496L}
{Line}, M.~R., {Liang}, M.~C., \& {Yung}, Y.~L. 2010, \apj, 717, 496

\bibitem[Masuda(2015)]{2015ApJ...805...28M} Masuda, K.\ 2015, \apj, 805, 28 

\bibitem[{{Mazeh} {et~al.}(2012){Mazeh}, {Nachmani}, {Sokol}, {Faigler}, \&
  {Zucker}}]{2012AandA...541A..56M}
{Mazeh}, T., {Nachmani}, G., {Sokol}, G., {Faigler}, S., \& {Zucker}, S. 2012,
  \aap, 541, A56

\bibitem[{{Mislis} \& {Hodgkin}(2012)}]{2012MNRAS.422.1512M}
{Mislis}, D., \& {Hodgkin}, S. 2012, \mnras, 422, 1512

\bibitem[{{Morris} {et~al.}(2013){Morris}, {Mandell}, \&
  {Deming}}]{2013ApJ...764L..22M}
{Morris}, B.~M., {Mandell}, A.~M., \& {Deming}, D. 2013, \apjl, 764, L22

\bibitem[{{Moses} {et~al.}(2011){Moses}, {Visscher}, {Fortney}, {Showman},
  {Lewis}, {Griffith}, {Klippenstein}, {Shabram}, {Friedson}, {Marley}, \&
  {Freedman}}]{2011ApJ...737...15M}
{Moses}, J.~I., {Visscher}, C., {Fortney}, J.~J., {et~al.} 2011, \apj, 737, 15

\bibitem[Mullally et al.(2015)]{2015ApJS..217...31M} Mullally, F., 
Coughlin, J.~L., Thompson, S.~E., et al.\ 2015, \apjs, 217, 31 

\bibitem[{{O'Donovan} {et~al.}(2006){O'Donovan}, {Charbonneau}, {Mandushev},
  {Dunham}, {Latham}, {Torres}, {Sozzetti}, {Brown}, {Trauger}, {Belmonte},
  {Rabus}, {Almenara}, {Alonso}, {Deeg}, {Esquerdo}, {Falco}, {Hillenbrand},
  {Roussanova}, {Stefanik}, \& {Winn}}]{2006ApJ...651L..61O}
{O'Donovan}, F.~T., {Charbonneau}, D., {Mandushev}, G., {et~al.} 2006, \apjl,
  651, L61

\bibitem[{{P{\'a}l} {et~al.}(2008){P{\'a}l}, {Bakos}, {Torres}, {Noyes},
  {Latham}, {Kov{\'a}cs}, {Marcy}, {Fischer}, {Butler}, {Sasselov}, {Sip{\H
  o}cz}, {Esquerdo}, {Kov{\'a}cs}, {Stefanik}, {L{\'a}z{\'a}r}, {Papp}, \&
  {S{\'a}ri}}]{2008ApJ...680.1450P}
{P{\'a}l}, A., {Bakos}, G.~{\'A}., {Torres}, G., {et~al.} 2008, \apj, 680, 1450

\bibitem[Petigura et al.(2013)]{2013PNAS..11019273P} Petigura, E.~A., 
Howard, A.~W., 
\& Marcy, G.~W.\ 2013, Proceedings of the National Academy of Science, 110, 19273 

\bibitem[{{Placek} {et~al.}(2014){Placek}, {Knuth}, \&
  {Angerhausen}}]{2014ApJ...795..112P}
{Placek}, B., {Knuth}, K.~H., \& {Angerhausen}, D. 2014, \apj, 795, 112

\bibitem[{{Quintana} {et~al.}(2013){Quintana}, {Rowe}, {Barclay}, {Howell},
  {Ciardi}, {Demory}, {Caldwell}, {Borucki}, {Christiansen}, {Jenkins},
  {Klaus}, {Fulton}, {Morris}, {Sanderfer}, {Shporer}, {Smith}, {Still}, \&
  {Thompson}}]{2013ApJ...767..137Q}
{Quintana}, E.~V., {Rowe}, J.~F., {Barclay}, T., {et~al.} 2013, \apj, 767, 137

\bibitem[{{Quintana} {et~al.}(2014){Quintana}, {Barclay}, {Raymond}, {Rowe},
  {Bolmont}, {Caldwell}, {Howell}, {Kane}, {Huber}, {Crepp}, {Lissauer},
  {Ciardi}, {Coughlin}, {Everett}, {Henze}, {Horch}, {Isaacson}, {Ford},
  {Adams}, {Still}, {Hunter}, {Quarles}, \& {Selsis}}]{2014Sci...344..277Q}
{Quintana}, E.~V., {Barclay}, T., {Raymond}, S.~N., {et~al.} 2014, Science,
  344, 277

\bibitem[{{Rauer} \& {Catala}(2013)}]{2013arXiv1310.0696R}
{Rauer}, H., \& {Catala}, C. a.~a. 2013, ArXiv e-prints, arXiv:1310.0696

\bibitem[{{Ricker} {et~al.}(2010){Ricker}, {Latham}, {Vanderspek}, {Ennico},
  {Bakos}, {Brown}, {Burgasser}, {Charbonneau}, {Clampin}, {Deming}, {Doty},
  {Dunham}, {Elliot}, {Holman}, {Ida}, {Jenkins}, {Jernigan}, {Kawai},
  {Laughlin}, {Lissauer}, {Martel}, {Sasselov}, {Schingler}, {Seager},
  {Torres}, {Udry}, {Villasenor}, {Winn}, \& {Worden}}]{2010AAS...21545006R}
{Ricker}, G.~R., {Latham}, D.~W., {Vanderspek}, R.~K., {et~al.} 2010, in
  Bulletin of the American Astronomical Society, Vol.~42, American Astronomical
  Society Meeting Abstracts 215, 450.06

\bibitem[{{Rogers} {et~al.}(2013){Rogers}, {L{\'o}pez-Morales}, {Apai}, \&
  {Adams}}]{2013ApJ...767...64R}
{Rogers}, J., {L{\'o}pez-Morales}, M., {Apai}, D., \& {Adams}, E. 2013, \apj,
  767, 64

\bibitem[{{Rowe} {et~al.}(2014){Rowe}, {Bryson}, {Marcy}, {Lissauer},
  {Jontof-Hutter}, {Mullally}, {Gilliland}, {Issacson}, {Ford}, {Howell},
  {Borucki}, {Haas}, {Huber}, {Steffen}, {Thompson}, {Quintana}, {Barclay},
  {Still}, {Fortney}, {Gautier}, {Hunter}, {Caldwell}, {Ciardi}, {Devore},
  {Cochran}, {Jenkins}, {Agol}, {Carter}, \& {Geary}}]{2014ApJ...784...45R}
{Rowe}, J.~F., {Bryson}, S.~T., {Marcy}, G.~W., {et~al.} 2014, \apj, 784, 45


\bibitem[Rowe et al.(2015)]{2015ApJS..217...16R} Rowe, J.~F., Coughlin, 
J.~L., Antoci, V., et al.\ 2015, \apjs, 217, 16 


\bibitem[{{Russell}(1916)}]{1916ApJ....43..173R}
{Russell}, H.~N. 1916, \apj, 43, 173

\bibitem[{{Sanchis-Ojeda} \& {Winn}(2011)}]{2011ApJ...743...61S}
{Sanchis-Ojeda}, R., \& {Winn}, J.~N. 2011, \apj, 743, 61

\bibitem[{{Sanchis-Ojeda} {et~al.}(2013){Sanchis-Ojeda}, {Winn}, {Marcy},
  {Howard}, {Isaacson}, {Johnson}, {Torres}, {Albrecht}, {Campante}, {Chaplin},
  {Davies}, {Lund}, {Carter}, {Dawson}, {Buchhave}, {Everett}, {Fischer},
  {Geary}, {Gilliland}, {Horch}, {Howell}, \& {Latham}}]{2013ApJ...775...54S}
{Sanchis-Ojeda}, R., {Winn}, J.~N., {Marcy}, G.~W., {et~al.} 2013, \apj, 775,
  54

\bibitem[{{Santerne} {et~al.}(2011{\natexlab{a}}){Santerne}, {Bonomo},
  {H{\'e}brard}, {Deleuil}, {Moutou}, {Almenara}, {Bouchy}, \&
  {D{\'{\i}}az}}]{2011AandA...536A..70S}
{Santerne}, A., {Bonomo}, A.~S., {H{\'e}brard}, G., {et~al.}
  2011{\natexlab{a}}, \aap, 536, A70

\bibitem[{{Santerne} {et~al.}(2011{\natexlab{b}}){Santerne}, {D{\'{\i}}az},
  {Bouchy}, {Deleuil}, {Moutou}, {H{\'e}brard}, {Eggenberger}, {Ehrenreich},
  {Gry}, \& {Udry}}]{2011AandA...528A..63S}
{Santerne}, A., {D{\'{\i}}az}, R.~F., {Bouchy}, F., {et~al.}
  2011{\natexlab{b}}, \aap, 528, A63

\bibitem[{{Santerne} {et~al.}(2012){Santerne}, {Moutou}, {Barros}, {Damiani},
  {D{\'{\i}}az}, {Almenara}, {Bonomo}, {Bouchy}, {Deleuil}, \&
  {H{\'e}brard}}]{2012AandA...544L..12S}
{Santerne}, A., {Moutou}, C., {Barros}, S.~C.~C., {et~al.} 2012, \aap, 544, L12

\bibitem[{{Shporer} {et~al.}(2010){Shporer}, {Kaplan}, {Steinfadt}, {Bildsten},
  {Howell}, \& {Mazeh}}]{2010ApJ...725L.200S}
{Shporer}, A., {Kaplan}, D.~L., {Steinfadt}, J.~D.~R., {et~al.} 2010, \apjl,
  725, L200

\bibitem[{{Shporer} {et~al.}(2011){Shporer}, {Jenkins}, {Rowe}, {Sanderfer},
  {Seader}, {Smith}, {Still}, {Thompson}, {Twicken}, \&
  {Welsh}}]{2011AJ....142..195S}
{Shporer}, A., {Jenkins}, J.~M., {Rowe}, J.~F., {et~al.} 2011, \aj, 142, 195

\bibitem[{{Shporer} {et~al.}(2014){Shporer}, {O'Rourke}, {Knutson},
  {Szab{\'o}}, {Zhao}, {Burrows}, {Fortney}, {Agol}, {Cowan}, {Desert},
  {Howard}, {Isaacson}, {Lewis}, {Showman}, \& {Todorov}}]{2014ApJ...788...92S}
{Shporer}, A., {O'Rourke}, J.~G., {Knutson}, H.~A., {et~al.} 2014, \apj, 788,
  92

\bibitem[{{Smith} {et~al.}(2012){Smith}, {Stumpe}, {Van Cleve}, {Jenkins},
  {Barclay}, {Fanelli}, {Girouard}, {Kolodziejczak}, {McCauliff}, {Morris}, \&
  {Twicken}}]{2012PASP..124.1000S}
{Smith}, J.~C., {Stumpe}, M.~C., {Van Cleve}, J.~E., {et~al.} 2012, \pasp, 124,
  1000

\bibitem[{{Southworth}(2011)}]{2011MNRAS.417.2166S}
{Southworth}, J. 2011, \mnras, 417, 2166

\bibitem[{{Still} \& {Barclay}(2012)}]{2012ascl.soft08004S}
{Still}, M., \& {Barclay}, T. 2012, {PyKE: Reduction and analysis of Kepler
  Simple Aperture Photometry data}, astrophysics Source Code Library,
  ascl:1208.004

\bibitem[{{Stumpe} {et~al.}(2012){Stumpe}, {Smith}, {Van Cleve}, {Twicken},
  {Barclay}, {Fanelli}, {Girouard}, {Jenkins}, {Kolodziejczak}, {McCauliff}, \&
  {Morris}}]{2012PASP..124..985S}
{Stumpe}, M.~C., {Smith}, J.~C., {Van Cleve}, J.~E., {et~al.} 2012, \pasp, 124,
  985

\bibitem[{{Szab{\'o}} {et~al.}(2012){Szab{\'o}}, {P{\'a}l}, {Derekas}, {Simon},
  {Szalai}, \& {Kiss}}]{2012MNRAS.421L.122S}
{Szab{\'o}}, G.~M., {P{\'a}l}, A., {Derekas}, A., {et~al.} 2012, \mnras, 421,
  L122

\bibitem[{{Visscher} \& {Moses}(2011)}]{2011ApJ...738...72V}
{Visscher}, C., \& {Moses}, J.~I. 2011, \apj, 738, 72













\end{thebibliography}
\end{document}